\documentstyle{mn}
\newif\ifAMStwofonts

\ifoldfss
  \ifCUPmtlplainloaded \else
    \NewTextAlphabet{textbfit} {cmbxti10} {}
    \NewTextAlphabet{textbfss} {cmssbx10} {}
    \NewMathAlphabet{mathbfit} {cmbxti10} {} 
    \NewMathAlphabet{mathbfss} {cmssbx10} {} 
  \fi
  \ifAMStwofonts
    \ifCUPmtlplainloaded \else
      \NewSymbolFont{upmath} {eurm10}
      \NewSymbolFont{AMSa} {msam10}
      \NewMathSymbol{\upi}     {0}{upmath}{19}
      \NewMathSymbol{\umu}     {0}{upmath}{16}
      \NewMathSymbol{\upartial}{0}{upmath}{40}
      \NewMathSymbol{\leqslant}{3}{AMSa}{36}
      \NewMathSymbol{\geqslant}{3}{AMSa}{3E}

      \let\leq=\leqslant 
       
    \fi
  \fi
\fi 

\ifnfssone
  \newmathalphabet{\mathit}
  \addtoversion{normal}{\mathit}{cmr}{m}{it}
  \addtoversion{bold}{\mathit}{cmr}{bx}{it}
  \newmathalphabet{\mathbfit} 
  \addtoversion{normal}{\mathbfit}{cmr}{bx}{it}
  \addtoversion{bold}{\mathbfit}{cmr}{bx}{it}
  \newmathalphabet{\mathbfss} 
  \addtoversion{normal}{\mathbfss}{cmss}{bx}{n}
  \addtoversion{bold}{\mathbfss}{cmss}{bx}{n}
  \ifAMStwofonts
    \ifCUPmtlplainloaded \else
      \UseAMStwoboldmath
      \makeatletter
      \new@mathgroup\upmath@group
      \define@mathgroup\mv@normal\upmath@group{eur}{m}{n}
      \define@mathgroup\mv@bold\upmath@group{eur}{b}{n}
      \edef\UPM{\hexnumber\upmath@group}
      \new@mathgroup\amsa@group
      \define@mathgroup\mv@normal\amsa@group{msa}{m}{n}
      \define@mathgroup\mv@bold\amsa@group{msa}{m}{n}
      \edef\AMSa{\hexnumber\amsa@group}
      \makeatother
      \mathchardef\upi="0\UPM19
      \mathchardef\umu="0\UPM16
      \mathchardef\upartial="0\UPM40
      \mathchardef\leqslant="3\AMSa36
      \mathchardef\geqslant="3\AMSa3E

      \let\leq=\leqslant 

    \fi
  \fi
\fi 

\ifnfsstwo
  \DeclareMathAlphabet{\mathbfit}{OT1}{cmr}{bx}{it}
  \SetMathAlphabet\mathbfit{bold}{OT1}{cmr}{bx}{it}
  \DeclareMathAlphabet{\mathbfss}{OT1}{cmss}{bx}{n}
  \SetMathAlphabet\mathbfss{bold}{OT1}{cmss}{bx}{n}
  \ifAMStwofonts
    \ifCUPmtlplainloaded \else
      \DeclareSymbolFont{UPM}{U}{eur}{m}{n}
      \SetSymbolFont{UPM}{bold}{U}{eur}{b}{n}
      \DeclareSymbolFont{AMSa}{U}{msa}{m}{n}
      \DeclareMathSymbol{\upi}{0}{UPM}{"19}
      \DeclareMathSymbol{\umu}{0}{UPM}{"16}
      \DeclareMathSymbol{\upartial}{0}{UPM}{"40}
      \DeclareMathSymbol{\leqslant}{3}{AMSa}{"36}
      \DeclareMathSymbol{\geqslant}{3}{AMSa}{"3E}

      \let\leq=\leqslant 

    \fi
  \fi
\fi 

\ifCUPmtlplainloaded \else
  \ifAMStwofonts \else 
    \def\upi{\pi}
    \def\umu{\mu}
    \def\upartial{\partial}
  \fi
\fi
\title{Gamma-rays from cascades in close massive binaries containing 
energetic pulsars}  
\author[A. Sierpowska \& W. Bednarek]
       {A. Sierpowska \& W. Bednarek \\
        Department of Experimental Physics, University of \L \'od\'z,
        ul. Pomorska 149/153, 90-236 \L \'od\'z, Poland}
\date{Accepted 
      Received ;
      in original form }
\pubyear{}
\begin{document}
\maketitle

\begin{abstract}
Some massive binaries should contain energetic pulsars which inject
relativistic leptons from their inner magnetospheres and/or pulsar wind
regions. If the binary system is compact enough, then these leptons can 
initiate inverse Compton (IC) $e^\pm$ pair cascades in the anisotropic radiation field 
of a massive star. $\gamma$-rays can be produced in the IC cascade
during its development in a pulsar wind region 
and above a shock in a massive star wind region where the propagation of leptons is
determined by the structure of a magnetic field around the massive star.
For a binary system with specific parametres, we calculate phase dependent spectra 
and fluxes 
of $\gamma$-rays escaping as a function of the inclination angle 
of the system and for different assumptions on injection conditions of the
primary leptons (their initial spectra and location of the shock inside the binary).
We conclude that the features of $\gamma$-ray emission from such massive binaries 
containing energetic pulsars should allow
to obtain important information on the acceleration of particles by the 
pulsars, and on interactions of a compact object with the massive star wind.
Predicted $\gamma$-ray light curves and spectra in the GeV and TeV energy ranges 
from such binary systems within our Galaxy and Magellanic Clouds 
should be observed by future AGILE and GLAST satellites and low threshold 
Cherenkov telescopes such as MAGIC, HESS, VERITAS or CANGAROO III.
\end{abstract}

\begin{keywords}
gamma-rays: massive binaries -- gamma-rays: theory -- radiation mechanisms: nonthermal
\end{keywords}

\section{Introduction}

It seems obvious that some close massive binary systems should contain 
young neutron 
stars able to accelerate leptons and possibly ions to relativistic energies. 
In fact, numerical simulations of the evolution of neutron stars show
that the fraction of systems with not accreting pulsars (during so called ejector phase) 
may become several percent of the total number of massive binaries (Lipunov~1990). 
However, due to their proximity to the high density winds of the massive stars
most of such systems do not show clear modulation of their radio signal with 
 pulsar rotational period. Therefore, only in the case of broad or highly eccentric
binaries the pulsars with periods significantly shorter than 1 second have been 
discovered (e.g. PSR B1259-63 with the period of 47.8 ms or A0538-66 with the 
period of 69.2 ms). 
TeV $\gamma$-ray emission, modulated with the period of $\sim 12.59$ ms, has been also
claimed from the compact binary Cyg X-3 (Brazier et al.~1990), but this result has not 
been confirmed by any other experiment. Other binaries are suspected 
to contain short period pulsars (e.g.  LSI +61 303, Maraschi \& 
Treves~1981), or stellar mass black holes ejecting relativistic particles highly 
anisotropically (e.g. Cyg X-1, Bednarek et al.~1990). 

In fact, observations of high energy X-ray and $\gamma$-ray sources to be coincident
with the locations of some binary systems support the hypothesis that the high energy 
processes play an important role in these sources. For example, EGRET 
sources have been found in directions towards e.g. LSI +61 303 (2EG J0241+6119, 
Thompson et al.~1995), Cyg X-3 (2EG J2033-4112, Mori et al.~1997), and 
LS 5039 (3EG J1824-1514, Paredes et al.~2000).
However, no TeV $\gamma$-rays from the above mentioned sources have been observed 
up to now (except early claims during the 80's which were not confirmed by subsequent 
more sensitive observations; see e.g. Weekes 1992). The limits on TeV emission from
PSR B1259-63, recently observed by the CANGAROO group at $\sim$47 days and at 
$\sim$157 days after the periastron (Kawachi et al.~2004), are above the 
theoretical predictions (Kirk et al.~1999, Kawachi et al.~2004, Murata et al.~2003). 
However, very recently the HESS group reported positive detection of this binary 
system on the level of $5\%$ of the Crab emission at energies above $\sim 400$ GeV 
with the power law spectrum and spectral index -2.8 (Schlenker et al.~2004). 
The TeV emission has been observed a few days before and after the periastron passage 
and declines towards the periastron moment.
Observations by the Whipple telescope of other four binary systems containing 
young pulsars 
(Hall et al.~2003) have not shown any positive detection of a steady or 
modulated TeV signal. 
Another binary system, Cen X-3, reported more recently in the TeV $\gamma$-rays 
(Chadwick et al.~1998, 1999, Atoyan et al.~2002), and also in the GeV $\gamma$-rays 
(Vestrand et al.~1997), contains slowly rotating and accreting neutron star.

In spite of not completely convincing high energy observational results, massive 
binaries are still often considered as likely sources of $\gamma$-rays produced mainly
in two general scenarios: (1) the anisotropic injection of particles 
from neutron stars or black holes, which interact with the 
radiation or matter inside the binary system (a massive companion, an accretion disk
and its corona or
a blob, see e.g. Cheng \& Ruderman~1991, Aharonian \& Atoyan~1991, 
Levinson \& Blandford~1996, Aharonian \& Atoyan~1999, Romero et al.~2001,
Atoyan et al.~2002, Georganopoulos et al.~2002, Romero et al.~2002, 
Romero et al.~2003, Bosch-Ramon \& Paredes~2004a,b, Orellana \& Romero~2004), 
or (2) the interaction of particles accelerated by a compact object or 
a shock wave created in collisions of the pulsar and stellar winds (e.g. 
Vestrand \& Eichler~1982, Harding \& 
Gaisser~1990, Tavani et al.~1994, Tavani \& Arons~1997, Kirk et al.~1999, 
Ball \& Kirk~2000, Murata et al.~2003) or two stellar 
winds (e.g. Eichler \& Usov~1993, Benaglia \& Romero~2003). 
In fact, the injection of relativistic particles and high energy
$\gamma$-rays in dense radiation field expected close to the accretion disks 
around compact objects (Carraminana~1992, Bednarek~1993) and massive companion stars 
(Protheroe \& Stanev~1987, Moskalenko et al.~1993) should result in copious 
$\gamma$-ray production. In very compact binaries and compact objects surrounded by 
luminous accretion disks, the primary particles can initiate IC $e^\pm$ pair 
cascades triggered due to the large optical depths. The $\gamma$-ray spectra  
are then produced with the characteristic cut-offs at a few 
tens of GeV (Bednarek~1997, Bednarek~2000). 
It was assumed in the latter works that 
the secondary $e^\pm$ pairs produced in the cascade process are isotropised inside the 
binary system by the random component of the magnetic field. 

In the present paper we consider in detail a more specific scenario in which the volume 
of the binary system is separated by the shock wave into two regions with different
properties. The shock appears as a result of collisions of the pulsar and massive star 
winds. We analyze very compact binary systems which contain
a young pulsar and a massive companion star of the OB or WR type able to create
soft radiation field in which the optical depths for relativistic leptons are much 
larger than unity. In fact, young pulsars in binary systems can be responsible for
quite different phenomena such as the appearance of jets called microquasars
(e.g Cyg X-3), binary radio pulsars (e.g. PSR 1259-63, SAX J0635+0533)
or a long period accreting objects (e.g. A0535+26). 
It is very difficult to observe some of such systems due to dense stellar winds 
produced by early type massive stars. However, binaries hidden dense winds 
should manifest themselves by the 
presence of large nonthermal luminosities which can, as we argue below, peak in 
the $\gamma$-ray energies. In Sect.~2 we define general conditions inside
the massive binary in which the termination shock is created in collisions of the pulsar and 
stellar winds. The injection of relativistic leptons (electrons or $e^\pm$ 
pairs) by the pulsar, their subsequent propagation and interaction with the radiation 
of the massive star is considered. In Sect.~3 we calculate production of 
$\gamma$-rays in IC $e^\pm$ cascades initiated by these leptons in the pulsar wind 
zone (PWZ) by applying the Monte Carlo method. Because of partial absorption of 
these $\gamma$-rays in the massive star 
wind region (MSWR), the secondary $e^\pm$ pairs are created and initiate there the second
part of the cascade.
This cascade differ significantly from the cascade in the PWZ due to the 
presence of ordered magnetic fields. The spectral and angular features of $\gamma$-rays 
escaping from the PWZ and MSWR are discussed in Sect.~5. 
As an example we consider the binary system with the parameters
derived for the WR star in the Cyg X-3 binary in which there are some observational 
evidence of the existence of a very young and energetic pulsar with
the period of several milliseconds.

\section{A pulsar close to a massive star}

We consider the binary system containing a fast pulsar and a massive 
companion star ($M \approx 10 M_{\odot}$). It is assumed that the energy loss rate of the 
pulsar, $L_{\rm rot}$, is high enough that the matter from the massive companion
can not accrete onto the pulsar surface neither from the outflow through the
Lagrangian point or from the dense stellar wind. This means that the binary system
has to be in the ejector phase, in which the structure of the inner pulsar 
magnetosphere, where efficient acceleration and $e^\pm$ bf{pair creation occur}, is not 
destroyed. The binary systems having the pulsar with the period fulfilling 
the condition (Harding \& Gaisser~1990), 
\begin{eqnarray}
P_{\rm ms} < 31 B_{12}^{4/7}{\dot M}_{18}^{-2/7}, 
\label{eq1}
\end{eqnarray}
should blow away accreting material, where $P_{\rm ms}$ is the pulsar period in ms,
$B = 10^{12}B_{12}$ G is its surface magnetic field, and ${\dot M} = 
10^{18}{\dot M}_{18}$ g s$^{-1}$ is the accretion rate.

Let us assume that the pulsar is on a circular orbit around the massive star
with the radius $R_{\rm s}$, effective surface temperature $T_{\rm s}$, and 
surface magnetic field $B_{\rm s}$. The star creates the wind which has 
the termination velocity $v_{\infty}$ and is characterized by the mass loss rate 
$\dot{\rm M}$. The separation of the stars is $D$. As a result of the interaction
of the pulsar and stellar winds a double shock structure is formed, separated by the 
contact discontinuity, at the distance determined by the above mentioned 
parameters of the stars. We apply the simplified model for the structure of the 
colliding winds based on momentum conservation (Girard \& Wilson 1987).
In this model the shock structure reaches a steady state configuration under 
the assumption of negligible pulsar orbital motion. 
In this case, the shock geometry of specific binary system 
is determined by the quantity $\eta$ (Ball \& Dodd, 2000), defined as follows:
\begin{equation}
\eta =  L_{\rm rot} /(\dot{M} V_{\rm w}c),
\label{eq2}
\end{equation}
\noindent
where the wind velocity, $V_{\rm w}$, depends on the radial distance, $r$, from the 
massive star according to (Hamann~1985)
\begin{equation}
V_{\rm w}(r) = v_\infty(1 - R_{\rm A}/r)^{1.5},
\label{eq3}
\end{equation}
\noindent
where $R_{\rm A}$ is the Alfven radius which can be derived by solving the equation  
$(1-R_{\rm s}/R_{\rm A})=\xi (R_{\rm s}/R_{\rm A})^4$. It has simple approximate 
solution 
\begin{eqnarray}
R_{\rm A} = R_{\rm s} \times \left\{ \begin {array}{ll}
1+\xi , & \xi \ll 1 \\
{\xi}^{1/4}, & \xi \gg 1, \end{array} \right.
\label{eq4}
\end{eqnarray}
\noindent
where $\xi = {B_{\rm s}}^2R_{\rm s}/(\dot{M} v_{\infty})$.

\noindent
The closest radial distance from the pulsar to the termination shock (measured in the plane of the
binary system) is then equal to,
\begin{equation}
\rho_{0} =  D \ \sqrt{\eta}/(1+\sqrt{\eta}).
\label{eq5}
\end{equation}
\noindent
In order to estimate $\rho_0$ we have to solve the set of Eqs.~\ref{eq2}-\ref{eq5}.
Note, that for $\eta < 1$ the star wind dominates over the pulsar wind and the termination shock 
wraps around the pulsar. For $\eta > 1$ the shock wraps around the companion star.
For large distances from the massive star the shock tends asymptotically to a cone 
characterized by a half-opening angle $\psi$,
\begin{equation}
\psi = 2.1(1-\eta^{-2/5}/4) \eta^{-1/3}.
\label{eq6}
\end{equation}
\noindent
We approximate the structure of the termination shock by the sphere of radius  
$\rho_{0}$ for $\theta > \theta_{0}$ ($\theta$ is defined in Fig.~1), and by a cone for 
$0 < \theta < \theta_{0}$, where the angle $\theta_{0}$ is defined as $\theta$ and is
given by the perpendicular line from pulsar to the generator of the cone. 
The shock is unterminated for $\theta < \psi$. 

\begin{figure}
  \vspace{6.truecm}
  \includegraphics{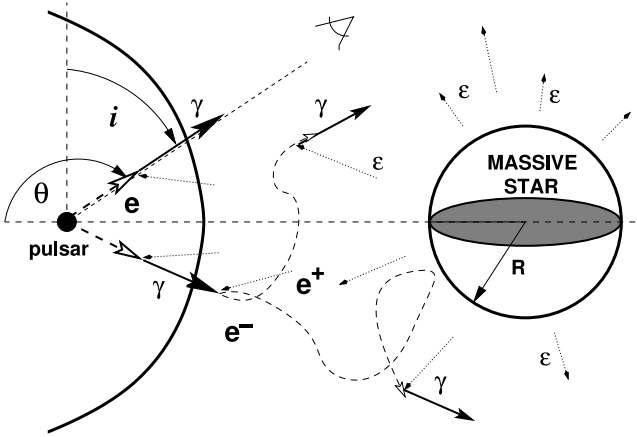}
  \caption{The schematic scenario of the interacting neutron star and massive companion
inside the compact binary system. Primary leptons, with the Lorentz factors $\gamma_e$, 
are injected by the pulsar and propagate inside the pulsar wind region 
(PWZ) along direction defined by the angle $\theta$, 
comptonizing soft radiation from the massive star. 
Secondary $\gamma$-rays can be absorbed in the same radiation field either inside 
the PWZ or, after passing the pulsar wind termination shock, in the massive star wind 
region (MSWR), triggering an electromagnetic cascade. 
Leptons propagate radially inside the PWZ but follow the 
magnetic field structure inside the MSWR. $\gamma$-rays from the cascade either escape from 
the binary system or fall onto the surface of the massive star, depending
on the injection parameters of the primary leptons (their energies and directions) and 
the parameters of the binary system (the massive star parameters and system separation).} 
\label{fig1}
\end{figure}

It is assumed in these simple estimations that the pulsar and stellar winds
are symmetric. In a more realistic case the dependence of the wind pressure on the 
distance from the equatorial plane of both stars should be taken into account. 
This might result in more complicated structure of the shock surface and 
involve more parameters which would prevent clear analysis of the angular features of 
high energy $\gamma$-ray emission from such compact binaries.
In a future work (Sierpowska \& Bednarek~2004a), in which we apply such 
a model to the $\gamma$-ray production in the binary pulsar PSR 1259-63 on the orbit
around the Be star, the effects associated with the non-spherical stellar wind will be 
taken into account.

In order to perform detailed calculations in the case of specific binary we apply as an 
example the parameters expected for the massive WR star in the Cyg X-3 system. 
This is a short period 
compact binary, $\tau = 4.8$ hr, with the massive star radius $< 3-6 R_{\odot}$,
surface temperature $T_{\rm s} > (7-9)\times 10^4$ K,
separation of the components $3.2 < D/R_{\odot} < 5.6$, and  the mass loss rate  
$\dot{M} \sim 1.1 \times 10^{-5} M_{\odot}\ yr^{-1}$ (Cherepashchuk \& Moffat~1994).
The orbital, phased resolved infrared spectra in the outburst 
and quiescent stages are consistent with the high orbital inclination of the system
with respect to the observer at the Earth, $i > 60^{\rm o}$, provided that the mass 
of the massive star is in the range, $5 < M_{WR}/M_{\odot} < 11$ (Hanson et al.~2000).
The infrared observations by van Kerkwijk et al.(2002) are consistent with the high 
orbital inclination, $i = 74^{\rm o}$, and the mass loss rate in the range 
$\dot{M} \sim 1.2 \times 10^{-4} M_{\odot}$ yr$^{-1}$ (based on the infrared data) and
$\dot{M} \sim 0.6 \times 10^{-5} M_{\odot}$ yr$^{-1}$ (based on the increase of orbital 
period). The stellar wind has terminal velocity $V_{\infty} = 1.45 \times 10^3$ km s$^{-1}$.
On the other hand, the inclination angle of system derived from the X-ray 
Chandra data is much lower, $i \approx 24^{\rm o}$. The mass of the compact object 
has been constrained by $< 3.6 M_{\odot}$, and the mass and radius of
the stellar companion  by $< 7.3 M_{\odot}$ and 
$R_{\rm s} < 1.6 R_{\odot}$ (Stark \& Saia~2003).

Based on the above observational constraints, we adopt the following parameters for our 
specific binary system.
A Wolf-Rayet type star with a radius $R_{\rm s}=1.6 \times R_{\odot}$, 
an effective temperature, $T_{\rm eff}=1.36\times 10^5$K, and a typical surface magnetic 
field, $B_{\rm s} \sim 10^2 - 10^3$G (Usov \& Melrose). The mass loss rate is 
$\dot{M} \sim 0.8 - 8.0 \times 10^{-5} \ M_{\odot}$ yr$^{-1}$, the velocity of the stellar wind 
at infinity $v_{\infty} \sim (1-5)\times 10^8$ cm s$^{-1}$, and the star rotational velocity 
$v_{rot} \sim (0.1 - 0.2) v_{\infty} $. 
The pulsar has a period $P_{\rm ms} = 12.59$
and a surface magnetic field $B_{\rm s} = 4.95\times 10^{11}$ G (Brazier et al.~1990). 
It is on a circular orbit, with a separation of $D = (3.6\pm 1.2)R_{\odot} = 2.25 \times 
R_{\rm s}$ (Stark \& Saia~2003). For these parameters, the pulsar energy loss rate is,   
\begin{eqnarray}
L_{\rm rot}\approx 6\times 10^{43}\ B_{12}^2\ P_{\rm ms}^{-4}~{\rm erg}~{\rm s}^{-1}\approx 
6\times 10^{38}~{\rm erg}~{\rm s}^{-1}. 
\label{eq7}
\end{eqnarray}
\noindent
For the above mentioned parameters of the massive 
companion in the Cyg X-3, the value of the parameter $\eta$ (see Eq.~\ref{eq2})  
is in the range $0.067 < \eta < 0.67$ (for fixed $V_{\infty} = 1.45 \times 10^3$ 
km s$^{-1}$). For further calculations we apply the average value $\eta = 0.3$
and for comparison the lowest value $\eta = 0.06$.

The conditions in the two regions of the binary system, i.e the PWZ and the MSWR,
separated by the shock 
structure differ significantly. Below the shock, inside the PWZ, relativistic 
leptons are frozen in the magnetized pulsar wind which propagate 
radially from the pulsar. 
Therefore, their synchrotron losses are neglected in our cascade calculations. 
We consider the IC cascade in the PWZ assuming that the thermal radiation from
the massive star dominates in the main volume of this region, neglecting the other
sources of soft photons e.g. such as the thermal emission from the neutron star surface 
or the nonthermal emission from the inner pulsar magnetoshere. This is true
if the temperature of the whole neutron star surface is below $T_{\rm NS}\sim 6\times 
10^6$ K. But such surface temperature is characteristic for the cooling   
neutron star in the age of only a few months after formation (Nomoto \& Tsuruta~1987).
This limit temperature has been obtained by comparing the energy density of 
thermal radiation from the massive star, 
$U_{\rm s}\approx \sigma_{\rm SB} T_{\rm s}^4(R_{\rm s}/D)^2$, 
where $\sigma_{\rm SB}$ is the Stefan-Boltzman constant, with the energy density
of thermal radiation from the neutron star, 
$U_{\rm NS}\approx \sigma_{\rm SB} T_{\rm NS}^4(R_{\rm NS}/R_{\rm LC})^2(1+\cos\theta)$,
at the distance of the light cylinder radius, $R_{\rm LC}$.
For the period of the pulsar as suggested 
in the Cyg X-3 binary, $P = 12.59$ ms, we obtain $R_{\rm NS}/R_{\rm LC}\approx 60$. 
The approximate angle of 
interaction between leptons and photons coming from the neutron star surface 
is then $\theta = \pi - R_{\rm NS}/R_{\rm LC}$ radians, assuming the case of a spherical 
wind. The estimate of the possible contribution of the nonthermal radiation from the 
pulsar inner magnetosphere is very difficult since the geometry of this
nonthermal emission is not well known. We comment that this nonthermal photons
has to be strongly collimated along the direction of motion of relativistic 
leptons as postulated by the observation of very narrow peaks in the light curves of 
young pulsars. Therefore, its importance should be strongly suppressed.
However, Bogovalov \& Aharonian~(2000) calculated possible $\gamma$-ray
emission from the vicinity of the light cylinder in the case of the Crab pulsar applying
a specific model for the radiation close to the light cylinder radius. 
According to these calculations the optical depth for leptons in the thermal radiation 
field of the Crab pulsar is very low at the light cylinder. However, in the nonthermal
radiation from the pulsar inner magnetosphere, the optical depth for leptons can be 
significant at the distance up to a 
few light cylinder radii. We conclude that the soft radiation in the main part of 
the PWZ is dominated by the thermal radiation from the massive star. 
Although, the important contribution of the nonthermal radiation from the inner pulsar 
magnetosphere at distances just above the light cylinder radius can not be excluded. 
This problem should be 
studied in a more detail when reliable models of the high-energy processes in the 
pulsar magnetospheres are better constrained.

Leptons which move through the PWZ interact efficiently 
with the soft radiation of the massive companion initiating IC $e^\pm$ pair
cascade. The charged products of this cascade arrive finally to the shock region in the 
pulsar wind and follow the flow along the shock surface. The power in these secondary
leptons is relatively low with respect to the power in secondary cascade $\gamma$-rays in 
the case of very close binary systems considered in this paper. 
The secondary $\gamma$-rays move into the massive star wind region. 
Some of them escape the binary system but a significant part can be converted into the 
next generation of $e^\pm$ pairs which have to follow the complex structure of the 
magnetic field present in the stellar wind.
These pairs can trigger further cascading processes producing 
next generation of $\gamma$-rays at directions which  
depend not only on the injection geometry of primary leptons but also on the geometry of 
the magnetic field. All these processes are discussed in detail in this paper  
and the spectra of $\gamma$-rays escaping at different angles 
with respect to the plane of the system and as a function of the phase of the pulsar on 
its orbit around the massive star are calculated.  

\subsection{Injection of leptons by the pulsar}

Young pulsars are efficient sources of energetic leptons, i.e. electrons and
positrons. These leptons originate in the cascade processes occurring in 
the inner pulsar magnetosphere, as expected in terms of the polar cap model 
(e.g. Ruderman \& Sutherland~1975, Arons \& Sharlemann~1979, Daugherty \& Harding~1982)
and the outer gap model (e.g. Cheng et al. 1986).
Leptons escape through the pulsar light cylinder to the pulsar 
wind zone (PWZ) where they can be additionally accelerated either just above 
$R_{\rm LC}$ (e.g. Beskin \& Rafikov~2000) or linearly through the PWZ 
(e.g. Contopoulos \& 
Kazanas 2002), or at the pulsar wind shock (e.g. Kennel \& Coronity 1984).

In this paper we consider two models for the primary spectra of leptons
injected into the radiation field of the massive star:

\begin{enumerate}

\item The power law spectrum of injected leptons between 100 MeV and 500 GeV and 
the spectral
index -1.2, as envisaged in the recent calculations of the spectra of leptons escaping 
from the inner magnetopshere performed by Hibschman \& Arons (2001). 

\item The monoenergetic injection of leptons with energies $10^6$ MeV corresponding 
to the Lorentz factors of the pulsar wind with the parameters typical for the
Crab pulsar. These leptons have very similar energies to those ones expected
for the supposed pulsar in Cyg X-3 binary system due to the similar value of $B/P^2$,
which determine the maximum potential drops through the pulsar magnetopshere.
In fact, adopting the formulas of the slot-gap pair-cascade model 
(Arons~1983) for a $e^\pm$ pair generation rate, $N_{e^\pm}$, and 
small ratio $\sigma_{\rm pw}$ of Poynting flux to kinetic energy flux in the wind
not far from the light cylinder radius, the Lorentz factor 
$\gamma_{\rm pw}$ of the pulsar wind can be estimated from (Melatos et al. 1995),

\begin{eqnarray}
\gamma_{\rm pw}  = L_{\rm rot}/(1+\sigma_{\rm pw})N_{e^\pm} m_{e^\pm}c^2 , 
\label{eq7}
\end{eqnarray}

\noindent
where $m_{e^\pm}$ is the electron mass, and $c$ is the velocity of light. 
For the considered here parameters of the pulsar 
$N_{e^\pm}\approx 2.8 \times 10^{37} s^{-1}$ and the pulsar wind Lorentz factor 
can be as high as $\gamma_{\rm pw} \sim 10^7$ (Melatos et al. 1995).
The small ratio $\sigma_{\rm pw}$ can be obtained in the narrow boundary layer near 
the light cylinder as suggested by Beskin \& Rafikov~(2000). 
These authors argue that the energy of the magnetized wind may be transfered to 
leptons with the Lorentz factors $\sim 10^6$.

\end{enumerate}

Most of the calculations presented below are shown for the power in 
injected spectrum of leptons normalized to $10^6$ MeV~sr$^{-1}$ in order to allow
direct comparison between the cases of power law and monoenergetic injection of 
leptons. The absolute fluxes of $\gamma$-rays produced in the 
cascade can be then obtained by multiplying by the factor $\sim 3\times 10^{37}$ 
s$^{-1}$ which is calculated from normalization of the power equal to
$10^6$ MeV to the total rotational energy lost by the pulsar (see Eq.~\ref{eq7}).   

\begin{figure}
\vspace{13.truecm}
\includegraphics{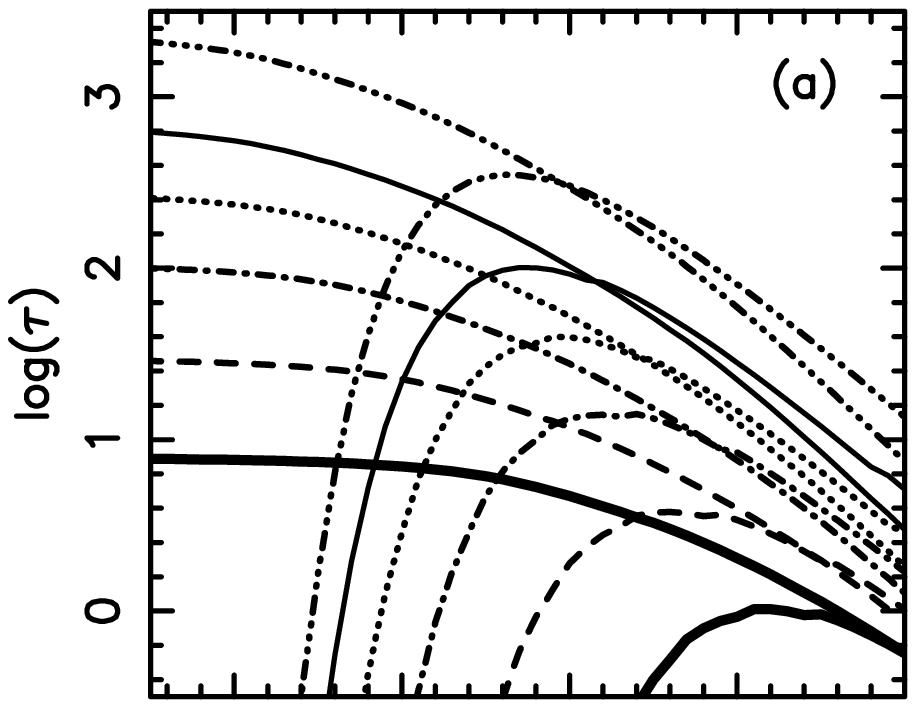}
\includegraphics{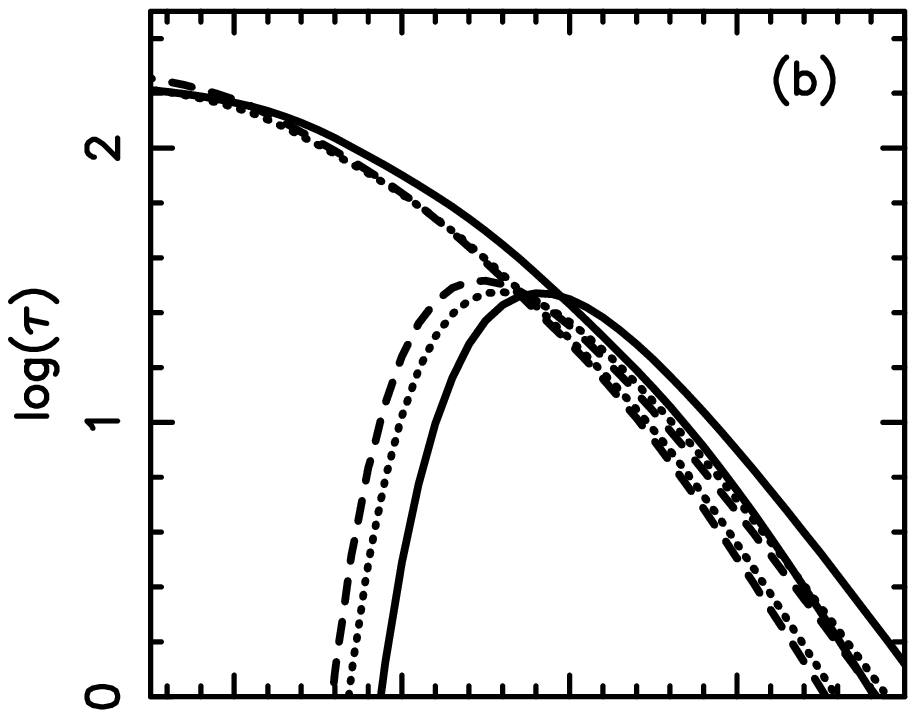}
\includegraphics{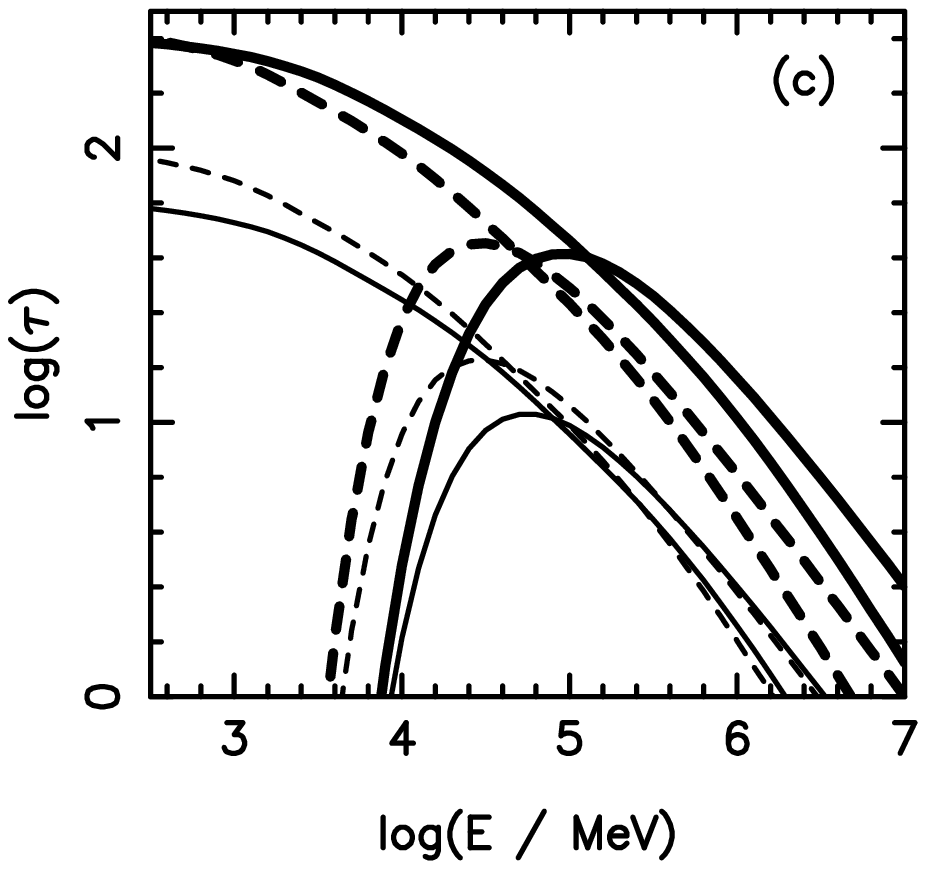}
\caption{The optical depths for leptons on ICS process, and for $\gamma$-rays on 
$e^\pm$ pair production in the anisotropic radiation of the massive star during 
their rectilinear propagation from the injection place at the distance 
$D = 2.25$R$_{\rm s}$ from the massive star up to the infinity (figure (a)).
The optical depths are shown as a function of particle energies for the case
of injection at the angle $\theta$, measured from direction defined by the centers of 
these stars (see Fig.~\ref{eq1}):  $\theta = 0^o$ (thick full curves), $30^o$ (dashed), 
$60^o$ (dot-dashed), $90^o$ (dotted), $120^o$ (thin full), and $150^o$ (
dot-dot-dot-dashed). (b) As in (a) but for particles propagating only to the termination 
shock defined by $\eta = 0.3$ and for the injection angles $90^o$ (solid lines),
$120^o$ (dotted lines) and $150^o$ (dashed lines). Note that for the 
angles $0^o - 60^o$, the optical depths are the same as in (a)
due to the lack of boundary on the rectilinear propagation caused by the presence of the 
shock. (c) As in (b) but for two specific locations of the shock defined by 
$\eta = 0.06$ and 0.6 and for the injection angles of primary leptons equal to $90^o$ 
(solid lines) and $150^o$ (dashed lines).} 
\label{fig2}
\end{figure}
\subsection{Conditions for propagation of leptons and gamma-rays}

\begin{figure*}
\vspace{8truecm}
\includegraphics{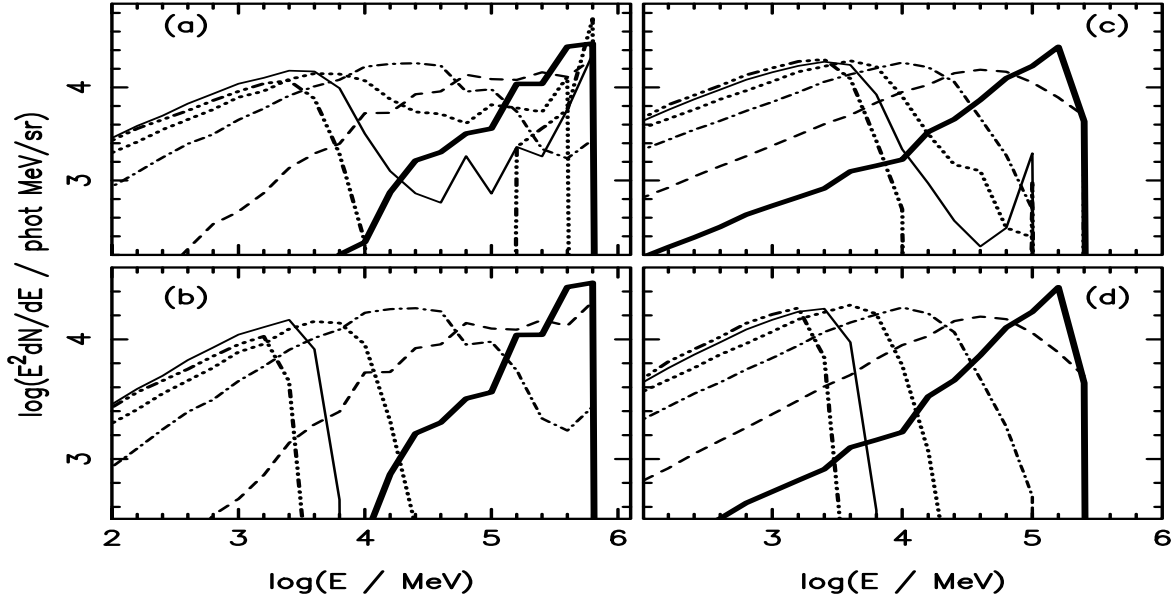}
\caption{The $\gamma$-ray spectra produced in the cascade inside the PWZ
in the case of injection of primary leptons with: the monoenergetic spectrum 
and energy $10^6$ MeV (on 
the left), and the power law spectrum (on the right). The spectra are shown 
for different injection angles of the primary leptons: $\theta = 0^o$ (thick solid curve), 
$30^o$ (dashed), $60^o$ (dot-dashed), $90^o$ (dotted), $120^o$ (thin solid), and
$150^o$ (dot-dot-dot-dashed). The $\gamma$-ray spectra which 
arrive to the shock region (defined by $\eta = 0.3$) are shown in (a) and (c), and 
the $\gamma$-ray spectra escaping to the observer at the infinity 
(after their absorption inside the MSWR) are shown in (b) and (d).
The spectra are obtained by sorting photons within the intervals with the width 
$\Delta$(log E) = 0.2 with the lower boundary of the interval marked in the figures.} 
\label{fig3}
\end{figure*}

We are interested in binary systems in which relativistic leptons, injected by 
the pulsar, can develop the IC $e^\pm$ pair cascades in the anisotropic radiation of the 
massive companion. Considered by us binary 
system meets this requirement if the optical depths
for IC scattering of massive star thermal radiation and for 
$\gamma$-rays absorption process in the same radiation are much larger than unity.
For the injection place of leptons and $\gamma$-rays equal to the separation distance 
between the pulsar and the WR star, $D = 2.25$R$_{\rm s}$, and other parameters
as mentioned above, we calculate the optical depths
up to the infinity for different angles of injection measured with respect to the 
direction defined by the centres of these stars and as a function of lepton energies 
(see Fig.~\ref{fig2}). The optical depths for leptons are very large in 
considered energy range. Moreover, $\gamma$-rays with
energies above a few GeV have high probability of interaction for all directions
(optical depth larger than unity). Therefore, leptons injected with energies large enough 
should develop IC $e^\pm$ pair cascade. 

In fact, the cascade can start to develop already inside the PWZ (see Fig.~\ref{eq2}b). 
The optical depths for all injection angles, calculated up to the location of the shock
defined by $\eta = 0.3$, are much larger than unity at specific 
energy ranges. In Fig.~\ref{eq2}c the dependence of the optical depths 
for leptons and $\gamma$-rays on the location of the shock inside the binary system is shown. 
It is clear that even for the extreme values of $\eta = 0.06$ and 0.6 leptons should 
develop efficient cascades if injected with energies above $\sim$10 GeV.  
>From comparison of Figs.~\ref{eq2}a and \ref{eq2}bc we find that the optical depths 
for $\gamma$-ray photons which are produced below the shock (in the PWZ) 
but propagate through the MSWR
are still much larger than unity. Therefore, these $\gamma$-rays partially convert into 
the next generation of $e^\pm$ pairs in the MSWR.
These $e^\pm$ pairs produce next generation of $\gamma$-rays which angular distribution 
on the sky may differ from initial injection directions of primary leptons due 
to the presence of a complex magnetic field in the MSWR. For that reason we consider the 
cascades developing inside the PWZ and MSWR separately.

\section{Pulsar wind region}

It is assumed that energetic leptons are injected from the vicinity of the 
pulsar light cylinder and propagate radially from the pulsar almost at rest
with respect to the pulsar wind. In this case we neglect the synchrotron losses
of leptons during their propagation up to the pulsar wind shock. If the optical 
depths are such as shown in Figs.~1, leptons develop IC $e^\pm$ pair cascade whose 
efficiency depends on the injection parameters of these primary particles 
(initial energies of leptons, their injection directions and parameters of the 
binary system and massive star). 
We assume that the cascade initiated by specific primary lepton in the
PWZ develops in one dimension, i.e. in the direction of propagation of the primary 
particle. Such one-dimensional cascade develops up to the 
pulsar wind shock. The secondary $\gamma$-rays
pass through the shock into the massive star wind region where they can be
additionally absorbed producing next generation of energetic $e^\pm$ pairs.
For considered large optical depths, only the $\gamma$-rays with energies below 
the threshold for $e^\pm$ pair production escape freely from the binary system.
The secondary $e^\pm$ pairs which are produced in the PWZ are captured by the magnetic 
field of the shock region and move along its surface with the pulsar wind plasma. 
However, in the case of compact binary systems such as Cyg X-3, 
we can neglect the contribution of secondary $e^\pm$ pairs to 
the total escaping $\gamma$-ray spectrum, because the part of 
energy of the primary leptons transfered to the secondary $e^\pm$ pairs is relatively 
low compared to that one transfered to the secondary $\gamma$-rays, 
as we show later. The influence of this part of leptonic 
cascade on the total $\gamma$-ray spectrum escaping from the system will be considered 
in the future work dedicated to wider binary systems.

\subsection{Gamma-ray spectra from the PWZ} 

We follow the development of such IC $e^\pm$ pair cascade in the anisotropic radiation
of the massive star assuming that the primary particles are injected in the place 
corresponding to the location of the pulsar in the binary system (the light cylinder 
radius can be safely neglected with respect to the characteristic dimension of the PWZ).
The procedure for the cascade Monte Carlo simulations in the PWZ is generally 
the same as used in Bednarek~(1997, 2000) except for the assumption on the local 
isotropisation of secondary $e^\pm$ pairs applied in that paper. In this paper it is 
assumed that the 
secondary cascade $e^\pm$ pairs follow the direction of the parent $\gamma$-rays. 
The distance of the first interaction of a lepton in the IC scattering process, 
$L_{\rm IC}$, is determined by sampling from the optical 
depth calculated for specific parameters of lepton injection place.
$L_{\rm IC}$ is found for the random number $P_1$ applying the formula, 

\begin{eqnarray}
P_1 = exp\left(-\int_{0}^{L_{\rm IC}}\lambda_{\rm IC}^{-1}(E,L,\theta) dL\right),
\label{eq8}
\end{eqnarray}

\noindent
where $\lambda_{\rm IC}(E,L,\theta)$ is the mean free path for IC scattering
of lepton with energy $E$, at the place defined by the propagation distance $L$ and 
the angle $\theta$ (see Fig.~1). $\lambda_{\rm IC}$ is calculated from the formula

\begin{eqnarray}
\lambda_{\rm IC}^{-1}(E,L,\theta) = \int d\mu(1+\mu)\int d\phi \int N(\varepsilon,\Omega)
\sigma_{\rm KN}d\varepsilon,
\label{eq8a}
\end{eqnarray}

\noindent
where $\mu = \cos\theta$ is the cosine of the angle between lepton and the incident 
soft photon, $N(\varepsilon, \Omega)$ is the differential density of soft photons, and 
$\sigma_{\rm KN}$ is the Klein-Nishina cross section. The limits of integrations in 
this formula are constrained by the geometry of the scattering process (propagation 
directions of leptons and thermal photons from the massive star) and the kinematics of 
the IC process. Since these limits are very complicated (several cases have to be
considered), they will be published separately in Sierpowska (2004). 
The distance at which secondary $e^\pm$ pair is produced in collisions of the $\gamma$-ray
photon with the soft photon is obtained in a similar way (see Eqs. \ref{eq8} and 
\ref{eq8a}) by replacing the Klein-Nishina cross section by the cross section for
$e^\pm$ pair production in the two photon collision (see Appendix A in Bednarek 1997). 
The subsequent places of the interaction of secondary leptons and $\gamma$-rays are 
determined in the same way by replacing 
the lower limit in the integral in Eq.~(\ref{eq8}) from the n-th cascade step 
by the upper limit in the integral from the previous, (n-1)-th cascade step.
The neutron star is relatively close to the massive star
with respect to its dimensions as the separation of the components is only 2.25 larger
than the radius of the massive star. Therefore, the geometrical effects connected with 
limited solid angle of impinging soft photons have to be taken into account 
when considering the IC scattering process. 

The energies of secondary particles
produced by parent leptons or $\gamma$-rays as a result of the interaction with 
the soft photons from the massive star are obtained for the random number $P_2$ 
by sampling from the differential energy distributions of these secondary particles
according to the relation 

\begin{eqnarray}
P_2 = \left(\int_0^{E_{\rm sec}}N dE\right)\left(\int_0^{E_{\rm sec, m}}N dE\right)^{-1},
\label{eq9}
\end{eqnarray}

\noindent
where $N$ is the differential spectrum of secondary 
particles ($e^\pm$ pairs or $\gamma$-rays), produced by the parent particle 
 with energy E($\gamma$-ray or lepton, respectively), calculated at a specific
location inside the PWZ which is defined by $L_{\rm IC}$ for the IC process and by 
$L_{\gamma-\gamma}$ for $\gamma$-ray absorption process (see for details the Appendix B 
and C in Bednarek 1997); 
$E_{\rm sec, m}$ is the maximum energy of produced secondary particle allowed
by the kinematics of specific process, i.e. IC scattering or $\gamma-\gamma$ absorption;
$E_{\rm sec}$ is the simulated energy of produced secondary particle. 
The energy of parent lepton after n-th interaction and the energy
of secondary leptons, produced in $\gamma-\gamma$ absorption, is found by applying  
energy conservation, i.e $E_{e^\pm, n} = E_{e^\pm, n-1} - E_{\gamma, n}$ and
$E_{\gamma, n-1} = E_{e^+, n} +E_{e^-, n}$. 
Since we are interested in high energy $\gamma$-rays, i.e $> 100 MeV$, the cascading 
procedure is switched off when the  
secondary leptons are cooled to energies below 500 MeV. Such leptons are not able to 
produce $\gamma$-rays above 100 MeV in the radiation described by the black body 
spectrum with temperature $\sim 10^5$ K, typical for the massive stars considered by us.
The simulation procedure of  the $\gamma$-ray photon interaction place 
and the energies of secondary leptons from photon absorption is analogous to what was described above.

\begin{quote}
\begin{table}

\begin{tabular}{ r | r | r | r | r | r | }
\hline
& &\multicolumn{2}{c}{ $\eta = 0.3$}&\multicolumn{2}{c}{ $\eta = 0.06$} \\
\hline
&$\theta$ & shock [\%]  & esc [\%] & shock [\%] & esc [\%] \\
\hline 
(a)&0&-&58&-&57\\
&30&-&70&-&70\\
&60&-&88&66&19\\
&90&84&53&62&10\\
&120&82&47&51&12\\
&150&83&31&50&14\\
&180&80&-&44&-\\
\hline 
(b)&0&-&59&-&62\\
&30&-&82&-&83\\
&60&-&94&84&60\\
&90&92&87&92&48\\
&120&90&74&84&52\\
&150&87&64&85&56\\
&180&87&-&85&-\\
\hline

\end{tabular}
\caption{
The percentage of primary leptons initial energy with monoenergetic (a) and 
power law spectra (b) (see text for details) which is transfered to the $\gamma$-rays in 
the cascade process for two locations of the shock defined by $\eta =0.3$ and $0.06$.
The part of energy in $\gamma$-rays which reached the shock region is marked by $shock$
and which menaged to escape to the infinity is marked by $esc$. For small angles $\theta$
the shock is not present, and for $\theta = 180^o$ $\gamma$-rays do not escape 
but collide with the surface of the massive star.}
\end{table}
\end{quote}
\begin{figure}
\vspace{6.3truecm}
\includegraphics{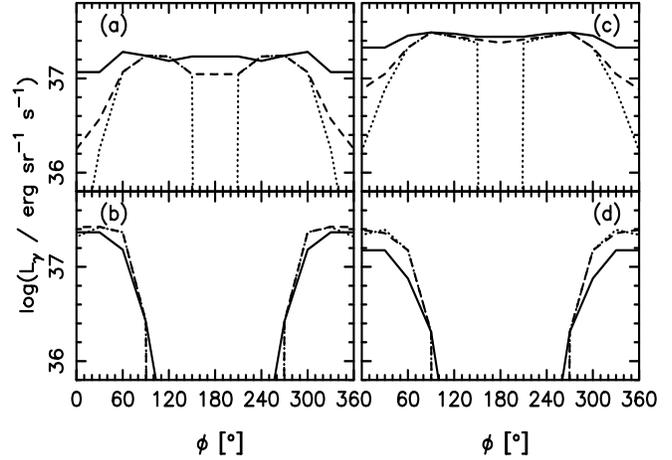}
\caption{The $\gamma$-ray light curves in two energy ranges,
$0.1 - 10$ GeV ((a) and (c)) and $10 - 10^3$ GeV ((b) and (d)),
produced in the cascade process occurring inside the PWZ. The cascade is 
initiated by leptons with the monoenergetic 
(figures on the left) and the power law spectra (on the right) for selected values 
of the inclination angles $i=30^o$ (solid curve), $60^o$ (dashed), and $90^o$ (dotted)
measured from the normal to the plane of the binary system.
} 
\label{fig4}
\end{figure}
\begin{figure*}
\vspace{8.5truecm}
\includegraphics{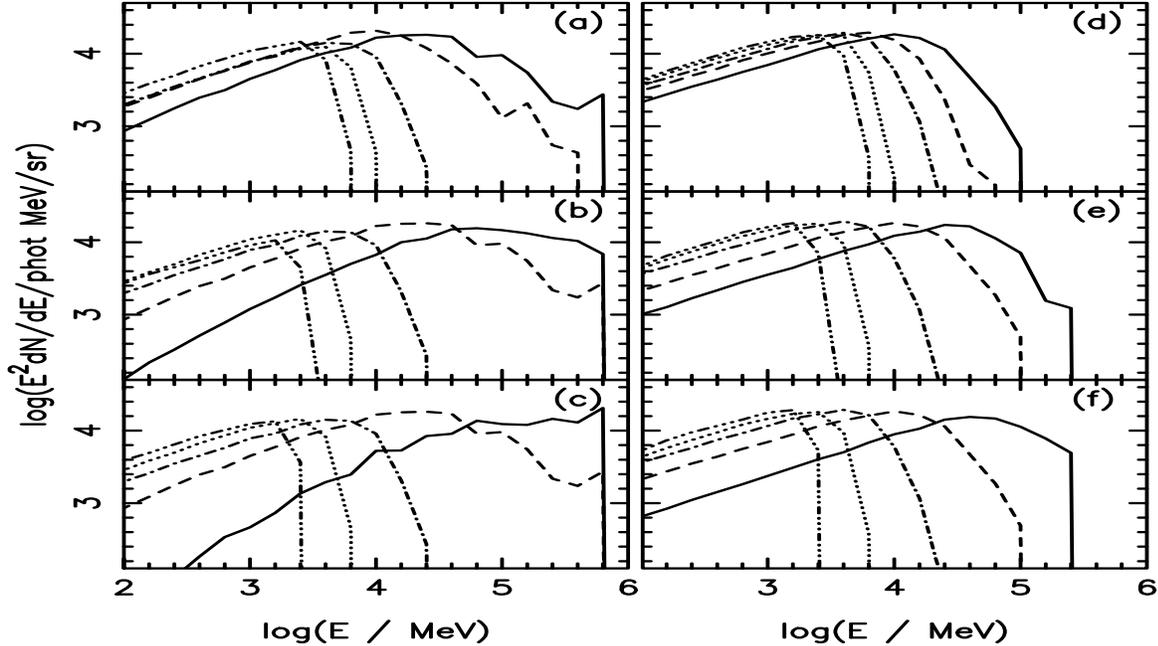}
\caption{The $\gamma$-ray spectra escaping from the binary system for different phases 
of the pulsar on its orbit around the massive star (measured with respect to the location 
of the observer), $\varphi = 30^o$ (solid curve) 
$60^o$ (dashed), $90^o$ (dot-dashed), $120^o$ (dotted), $180^o$ (dot-dot-dot-dashed),
and different inclination angles of the binary system
$i=30^o$ ((a) and (d)), $60^o$ ((b) and (e)), and $90^o$ ((c) and (f)). 
$\gamma$-rays are produced in the PWZ cascade by primary leptons with the 
monoenergetic (figures on the left) and the power law spectrum (on the right).} 
\label{fig5}
\end{figure*}

The cascade processes occurring inside the PWZ are axially symmetric with respect to 
the direction defined by the centers of stars. Therefore, the spectra of $\gamma$-rays, 
these at the location of the shock front and those ones escaping to the infinity through 
the radiation of the MSWR, are calculated only as a function 
of the injection angle $\theta$. The $\gamma$-ray spectra have been obtained for two, 
mentioned in Sect. 2.2, distributions of primary leptons and the location of the shock 
defined by $\eta = 0.3$ (see Figs.~3 for the monoenergetic and the power law 
spectra of primary leptons). 
Note that photons are sorted within the intervals with the width 
$\Delta(log E) = 0.2$. The basic features of the $\gamma$-ray spectra 
observed at infinity, i.e. after their partial absorption in the radiation field of the 
MSWR, can be easily understood if analyzed in the context of the optical depths shown in 
Figs.~1ab. Only primary leptons propagating in the outward direction (with respect to the 
massive star) can produce $\gamma$-ray fluxes above $\sim 100$ GeV, being potentially 
detectable by telescopes operating at very high energies (VHE) (Figs.~3bd). 
On the other hand, 
the $\gamma$-ray spectra escaping in the inward directions, i.e. close to the limb of the 
massive star, have comparable intensities at energies below 10 GeV (HE range)
within a factor of 2-3. Moreover, these spectra show cut-offs close to 
$\sim 10$ GeV which are determined by the surface temperature of the massive star. 
Therefore, the anticorrelation between the $\gamma$-ray emission in the HE and VHE ranges 
is expected from such compact luminous binaries.
The spectra of $\gamma$-rays produced in the PWZ and arriving to the shock region are 
shown in Figs.~3ac. The difference between $\gamma$-ray spectra
shown in Figs.~3ac and~3bd gives us the information about the importance of absorption 
processes in the MSWR. It is evident that the high energy part of the $\gamma$-ray 
spectrum produced in the PWZ is strongly absorbed during propagation through the MSWR if 
the primary 
leptons are injected in the hemisphere containing the massive star, i.e. for the angles 
$\theta >90^o$. Important information on the efficiency of $\gamma$-ray production 
can be derived from analysis of the percentage of the energy of primary leptons which 
is converted in the PWZ cascade process into the $\gamma$-ray photons. 
Table~1 reports these efficiencies at the infinity (marked by {\it esc}) and at 
the location of the shock inside the binary system (marked by {\it shock})  
defined by $\eta = 0.3$ and 0.06. Most of the initial lepton energy can be converted 
into the $\gamma$-rays in such a cascade already at the distance of the
shock in the case of dense soft radiation field of such compact massive companion.

A significant part of the power of produced $\gamma$-ray for the directions of primary 
leptons towards the massive star ($\theta > 90^o$) can be again converted into 
the next generation of $e^\pm$ pairs in the absorption process in the MSWR
(see the difference between $\it shock$ and $\it esc$ in Table~1). 
Therefore, possible contribution of the cascades initiated by these secondary
$e^\pm$ pairs in the MSWR to the total $\gamma$-ray spectrum escaping from the binary 
system can not be neglected. These secondary $e^\pm$ pairs 
propagate in the MSWR along the paths which are determined by the structure of the 
magnetic field in the wind of the massive star. That is why, the next generation of
$\gamma$-rays produced by them move in directions 
which can completely differ from the initial angular distribution of $\gamma$-rays 
escaping from the PWZ. This more complicated cascade with the re-distribution of 
charged particles with respect to their initial directions is discussed in details in the 
next section.

\subsection{Gamma-ray light curves and phase resolved spectra} 

In order to investigate in more details the angular dependence of the $\gamma$-ray
emission produced in the cascade inside the PWZ, we calculate the $\gamma$-ray light 
curves which should be observed at different inclination angles of the binary system. 
The results are shown for the two, HE and VHE, energy ranges (i.e. $0.1 - 10$ GeV and 
$10 - 10^3$ GeV) in the case
of primary leptons injected into the PWZ with the monoenergetic and power law 
distributions (see Figs. 4). The $\gamma$-ray luminosities are calculated after 
normalization of the power in primary leptons to the rotational energy lost by 
the pulsar, $L_e = L_{\rm rot}$. Clear anticorrelation
in the $\gamma$-ray curves is observed between these two energy ranges for both
initial spectra of primary leptons.
The VHE emission is mainly limited to the phases when the pulsar is in front of the
massive star ($\varphi = 0^o$). In contrast, the HE emission 
is much more uniform with significant decrease at phases around $0^o$ at which 
the VHE emission is the strongest. The disappearance 
of $\gamma$-ray emission for large inclination angles of $i > 90^o-\alpha$ (where 
$\alpha = 26.4^o$ is the angular extend of the massive companion observed from the 
distance of the binary separation) and at the phase close to $180^o$, is connected with 
the total eclipse of the pulsar by the massive companion.
The $\gamma$-ray spectra for selected phases of the pulsar on its orbit 
and the inclination angles mentioned in Figs. 4 are shown in Figs. 5. 
The $\gamma$-ray spectral shapes do not differ significantly below a few GeV 
(spectral index close to -1.5) for 
half of the pulsar phases, independently on the inclination angle of the binary
system. The level of this emission vary only by a factor of 2-3. In contrast, the VHE 
emission is limited to a relatively small range of phases with clear dependence 
on the inclination angle of the binary.

\section{Massive star wind region}

As we have shown above, a significant part of $\gamma$-rays from the PWZ 
which passed the termination shock is effectively absorbed in the 
MSWR by dense radiation field of the massive star 
(see Table.~1). However, the next generation of $e^\pm$ pairs is forced to 
follow the local magnetic field lines and their directions of propagation can 
change significantly with respect to the directions of their parent $\gamma$-rays. 
Therefore, the development of the cascade in the MSWR becomes much more complicated, 
determined by the complex magnetic field structure and the cascade is not further 
one-dimensional. The next generation $\gamma$-rays are usually produced at completely 
different angles than the initial directions of the PWZ cascade $\gamma$-rays.  

It is assumed in these cascade calculations that synchrotron energy losses of 
the cascading $e^\pm$ pairs can be neglected with respect to their energy losses on the ICS.
In fact, the comparison of energy densities of the magnetic field 
$\sim 2.5\times 10^{16}$ eV cm$^{-3}$, for the surface magnetic field of the star
$B_{\rm s} = 10^3$ G, and the thermal radiation $\sim 8.5\times 10^{17}$ eV cm$^{-3}$,
for the temperature of the star $T_{\rm s} = 1.36\times 10^5$ K, 
show that the IC losses 
should clearly dominate over the synchrotron process. Further from the star the energy 
density of radiation drops with the square 
of the distance and the energy density of magnetic field drops with the forth power
of the distance (see Eq.~(\ref{eq10}), note that the Alfven radius is very close to the 
stellar surface for the considered here parameters). Therefore, if the
IC losses dominate at the stellar surface it have to dominate everywhere above
the star. It has been shown in Bednarek (1997, see Eq. 4 in that paper) that 
the Bremsstrahlung energy losses 
of leptons in the wind of the massive star are negligible with respect to their IC losses
in the Thompson (T) regime for the parameters of the wind and the star considered here and
the Lorentz factors of leptons above the limiting the value $10^3$ considered in the 
cascade process. However in the present paper we also consider leptons with energies up 
to $10^6$ MeV. It is necessary to discuss in a more detail the case of the IC losses in 
the Klein-Nishina (KN) regime. The IC losses of leptons with the Lorentz factors $\gamma$
in the T regime and the thermal radiation with temperature $T_{\rm s}$ can be 
approximated by 

\begin{eqnarray}
\left({{dE}\over{dt}}\right) \approx 1.3\times 10^{-22} \gamma^2 T_{\rm s}^4 {~~~\rm MeV~s}^{-1}.
\label{eqA}
\end{eqnarray}
\noindent
We can estimate energy losses in the KN regime by 
putting into the above formula the Lorentz factor of leptons between 
the KN and T regimes, i.e. $\gamma \approx 2\times 10^4$ for $T_{\rm s} = 10^5$ K.
Then we get

\begin{eqnarray}
\left({{dE}\over{dt}}\right)_{\rm KN} \approx 5\times 10^{6} {~~~\rm MeV~s}^{-1}.
\label{eqB}
\end{eqnarray}
\noindent
Note that the logarithmic dependence of energy losses in the KN regime has been 
neglected. On the other hand, the bremsstrahlung energy losses can be approximated by

\begin{eqnarray}
\left({{dE}\over{dt}}\right)_{\rm brem} \approx 1.4\times 10^{-16} N E {~~~\rm MeV~s}^{-1}
\label{eqC}
\end{eqnarray}

\noindent
where N is the density of particles in cm$^{-3}$ and $E$ is energy of leptons in MeV.
For the extreme energies of injected leptons considered in the cascade process
($E = 10^6 MeV$), we obtain
   
\begin{eqnarray}
\left({{dE}\over{dt}}\right)_{\rm brem} \approx 1.4\times 10^{-10} N {~~~\rm MeV~s}^{-1}
\label{eqC}
\end{eqnarray}

\noindent
>From the comparison of $(dE/dt)_{\rm KN}$ with $(dE/dt)_{\rm brem}$,  
we obtain the critical density of the matter 
$N_{\rm cr} \approx 4\times 10^{16}$ cm$^{-3}$.

Let's estimate a typical density of the massive star wind in the MSWR close to
the stellar surface,

\begin{eqnarray}
N_{\rm w} = \dot{M}/4\pi R_{\rm s}^2V_{\rm w},
\end{eqnarray}

\noindent
which for the wind velocity $V_{\rm w} = 10^8$ cm s$^{-1}$ and the mass loss rate 
$\dot{M} = 4\times 10^{-5} M_\odot$ yr$^{-1}$, gives $\sim 8\times 10^{13}$ cm$^{-3}$.
So then, density of the massive star wind is 2-3 orders of magnitude lower 
than the critical density, $N_{\rm cr}$, above which the Bremsstrahlung energy losses  
dominate over the IC losses in the KN regime. Therefore, we conclude that the 
Bremsstrahlung 
process can be safely neglected with respect to the IC process for the model considered 
in this paper.

Below we describe the procedure applied for the calculation of the 
cascade processes with the presence of the magnetic field in the MSWR.

\subsection{Magnetic field structure inside MSWR}

The magnetic field in the wind of the massive star can have complicated structure.
In the region very close to the massive star surface it is characterized by dipolar 
component. At a certain distance, the radial component starts to dominate
due to the presence of the ionized plasma, and at larger distances the magnetic field  
becomes toroidal due to the rotation of the massive star. 
The strength of the magnetic field as a function of distance from the center of the massive star
can be described by the following equations (Usov \& Melrose, 1992),
\begin{eqnarray}
B(r) \approx B_{\rm s} \times \left\{ \begin {array}{ll}
(R_{\rm s}/r)^{3} , & R_{\rm s} \leq r < R_{\rm A} ,  \\
R_{\rm s}^3/(R_{\rm A} r^2) , & R_{\rm A} < r < R_{\rm tor} ,\\
(v_{\rm rot}/v_{\infty}) \ (R_{\rm s}^2/(R_{\rm A} r)) , & R_{\rm tor} < r ,  \end{array} \right.
\label{eq10}
\end{eqnarray}
\noindent
where R$_{\rm tor}$ is the radius of the toroidal 
field defined by the rotation velocity of the massive star $v_{\rm rot}$ and the star 
wind velocity at the infinity $v_{\infty}$, $R_{\rm tor} = R_{\rm s} 
v_{\infty}/v_{\rm rot}$.
In the case of the typical parameters of considered here for the WR star, the surface magnetic field is 
of the order of $B_{\rm s} = 10^3\ G$ and the $v_{\rm rot}\approx 0.1R_\infty$. The 
Alfven radius is located then at $R_{\rm A} = 1.12 \times R_{\rm s}$, and  
$R_{\rm tor} = 10\times R_{\rm s}$. Therefore, the main volume of the MSWR, in which 
IC $e^\pm$ pair cascade can efficiently develop, is dominated by the magnetic field with 
the radial structure.

\subsection{Propagation of leptons in MSWR}

The equations of motion of a charged particle in a constant uniform magnetic field 
(see Jackson~1962) are described in the coordinate system in which the vector of the
magnetic field, $\vec{B}$, is parallel to the axis OZ. Then, the coordinate of the
charged particle and its velocity is given by, 
\begin{eqnarray}
\vec{r}(t) &=& \vec{r}_0 + v_{\parallel}t \vec{e}_3 + a \cos \omega_B t \vec{e}_2 + a 
\sin \omega_B t \vec{e}_1,\\
\vec{v}(t) &=& v_{\parallel}t \vec{e}_3 + \omega_B a ( \cos \omega_B t \vec{e}_1 + \sin 
\omega_B t \vec{e}_2),
\end{eqnarray}
\noindent
where $\vec{e}_3$ is a unit vector parallel to the magnetic field line, $\vec{e}_1$ and 
$\vec{e}_2$ are 
the other orthogonal unit vectors, $\omega_B = e \vec{B} / \gamma m c = e c 
\vec{B} / E$ is the Larmor frequency and
$\omega_B a  = v_{\perp}$ is the perpendicular velocity of the charged particle 
with respect to the magnetic field line while $v_{\parallel}$  denotes component along 
the field, and $a$ is the Larmor radius.
Using these equations we follow the charged particle by determining its position and 
velocity vector with respect to the local magnetic field line. 

The step distance method is applied to follow the path of the particle in the magnetic 
field. The dimension of the step is chosen to be less than 0.01 of the Larmor radius in 
the local magnetic field. The optical depth for lepton on the IC scattering is calculated 
by summing up contributions from every step up to the moment of fulfilling the condition 
for simulating the place of interaction (given by Eq. \ref{eq8}). Note that this 
procedure is much more complicated and time consumming than applied in the PWZ since 
now we have to determine the direction of lepton at every step. 
We calculate the energy of produced secondary $\gamma$-ray photons applying the 
procedure described in Sect. 3.1 (see Eq. \ref{eq9}). The energy of 
surviving leptons is equal to the difference between its initial energy and the
energy of produced $\gamma$-ray photons. Therefore, after every scattering the 
conditions for propagation of leptons in the magnetic field change due to the change of
their Larmor radius. The particle paths are followed up to the moment
when leptons: (1) are cooled below 500 MeV, (2) move outside the sphere with the radius
of 10 massive star radii (since the efficiency of IC scattering at larger distances 
becomes low), (3) fall onto the massive star surface, or (4) move back through the shock plane.  
The energies and directions of produced $\gamma$-rays are stored and
used for analysis of their spectral and angular properties.

\begin{figure}
\vspace{13.8truecm}
\includegraphics{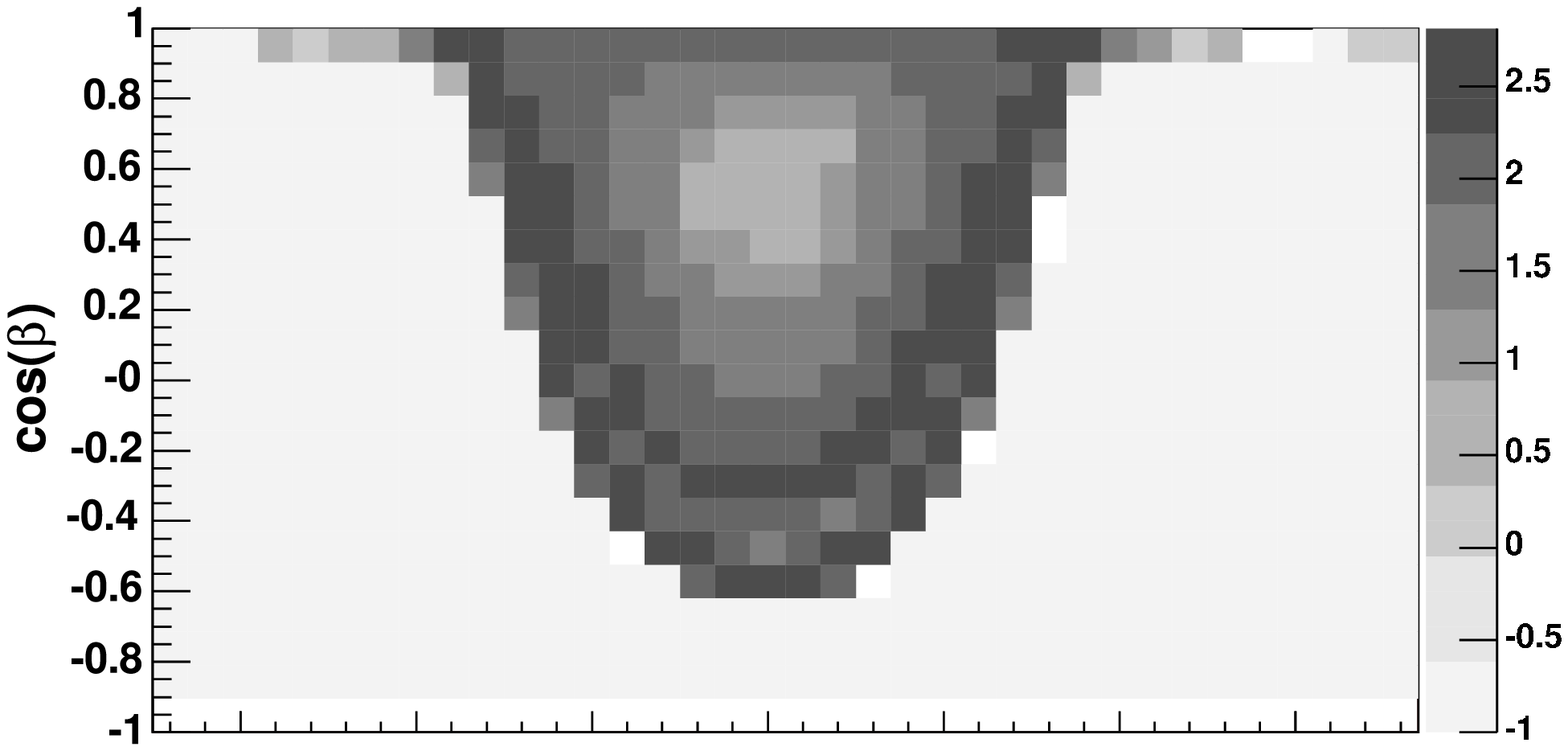}
\includegraphics{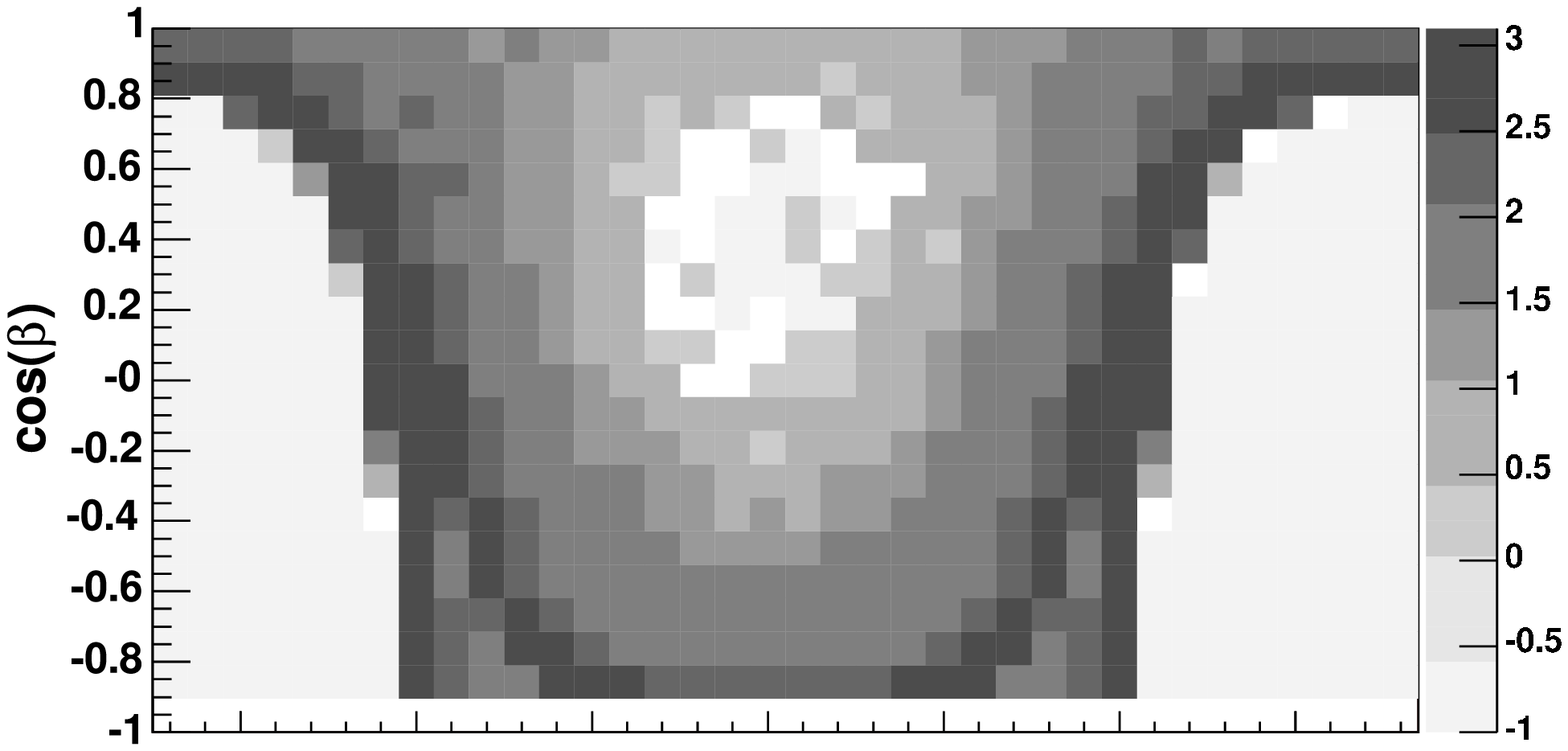}
\includegraphics{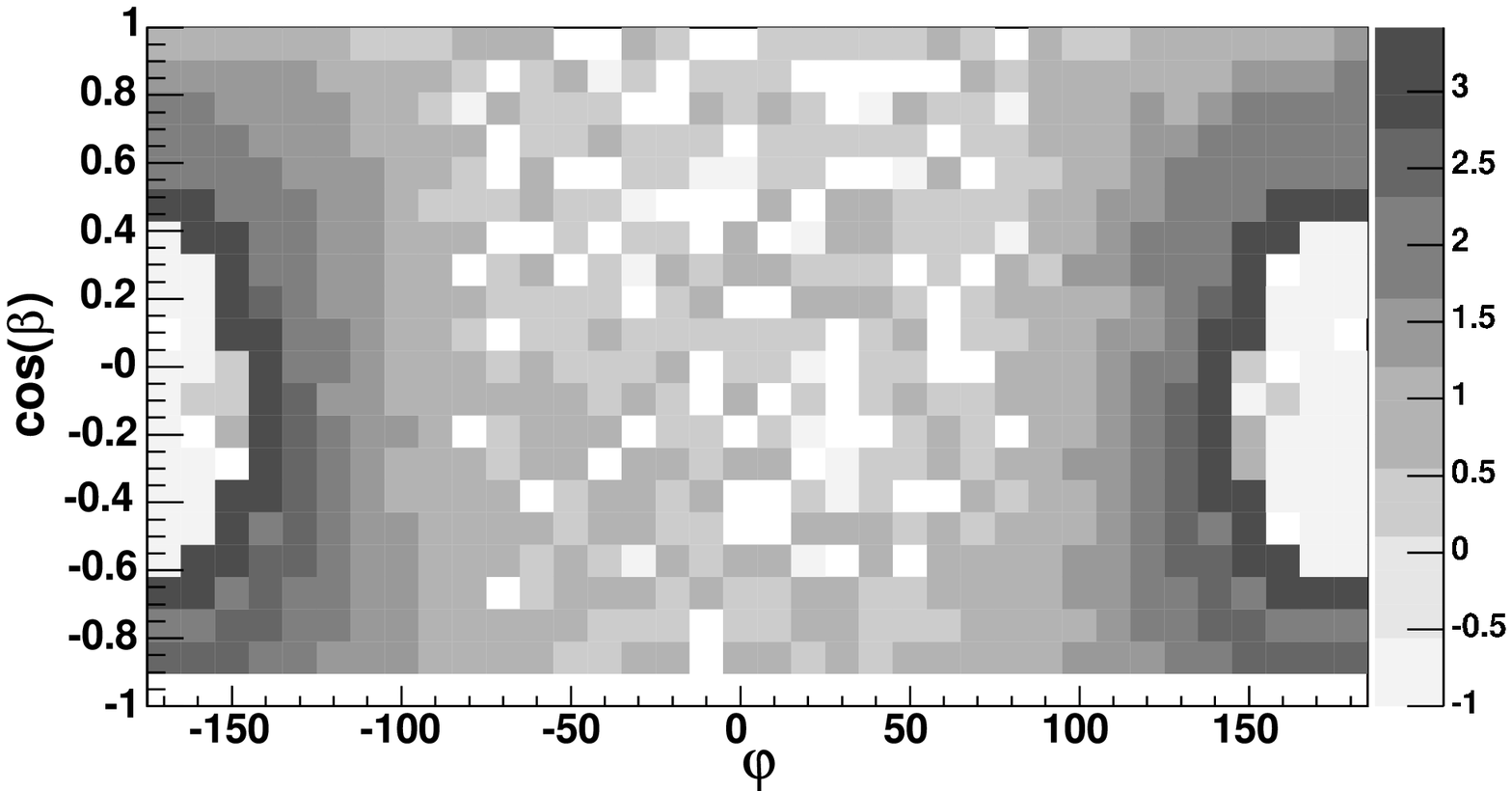}
\caption{The maps with the  numbers of $\gamma$-ray photons with energies $> 100$ MeV 
(in log grey scale)
escaping at specific directions on the sky defined by the azimuthal angle 
$\varphi$ and the cosine of the vertical angle $\beta$ (the bins with 
$\Delta\varphi = 10^{\rm o}$ and $\cos\beta = 0.1$).
$\varphi = 0^o$ correspond to the case of the massive star in front of 
the pulsar which is the source of primary leptons. These $\gamma$-rays are produced in 
the cascade in the MSWR initiated by the secondary cascade $\gamma$-rays escaping from 
the PWZ, which in turn were produced by primary monoenergetic leptons with energies 
$10^6$ MeV. The results, averaged over $N$ simulated primary leptons, are shown for 
different initial injection angles of the primary leptons
equal to $\theta = 90^o$ for $N = 200$ (a), $120^o$ for $N = 100$ (b), and $150^o$ for 
$N = 60$ (c), which are measured above 
the plane of the binary system (i.e. towards positive vertical angles).
Presented results are re-normalized for one lepton with energy $10^6$ MeV.}
\label{fig5a}
\end{figure}
\begin{figure}
\vspace{13.8truecm}
\includegraphics{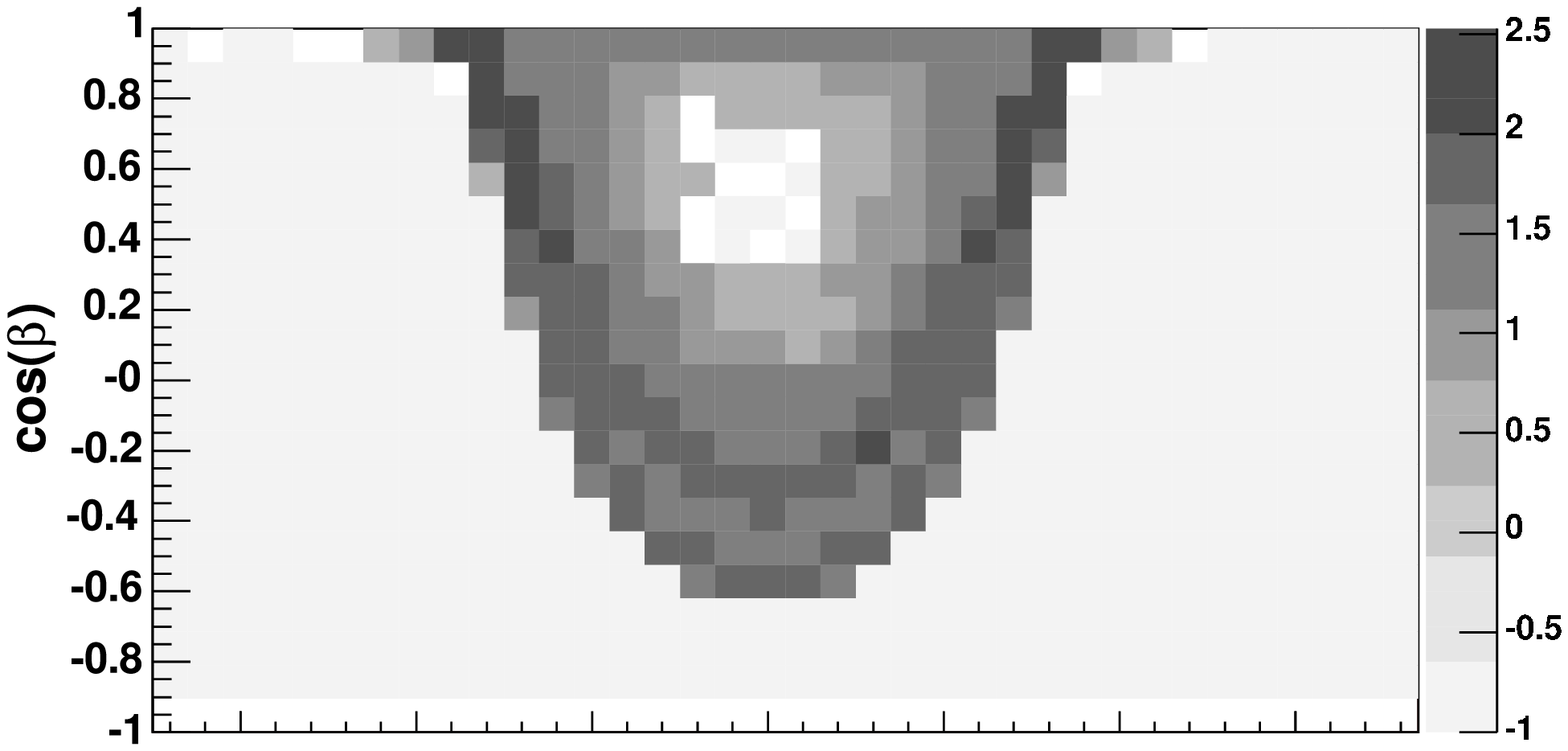}
\includegraphics{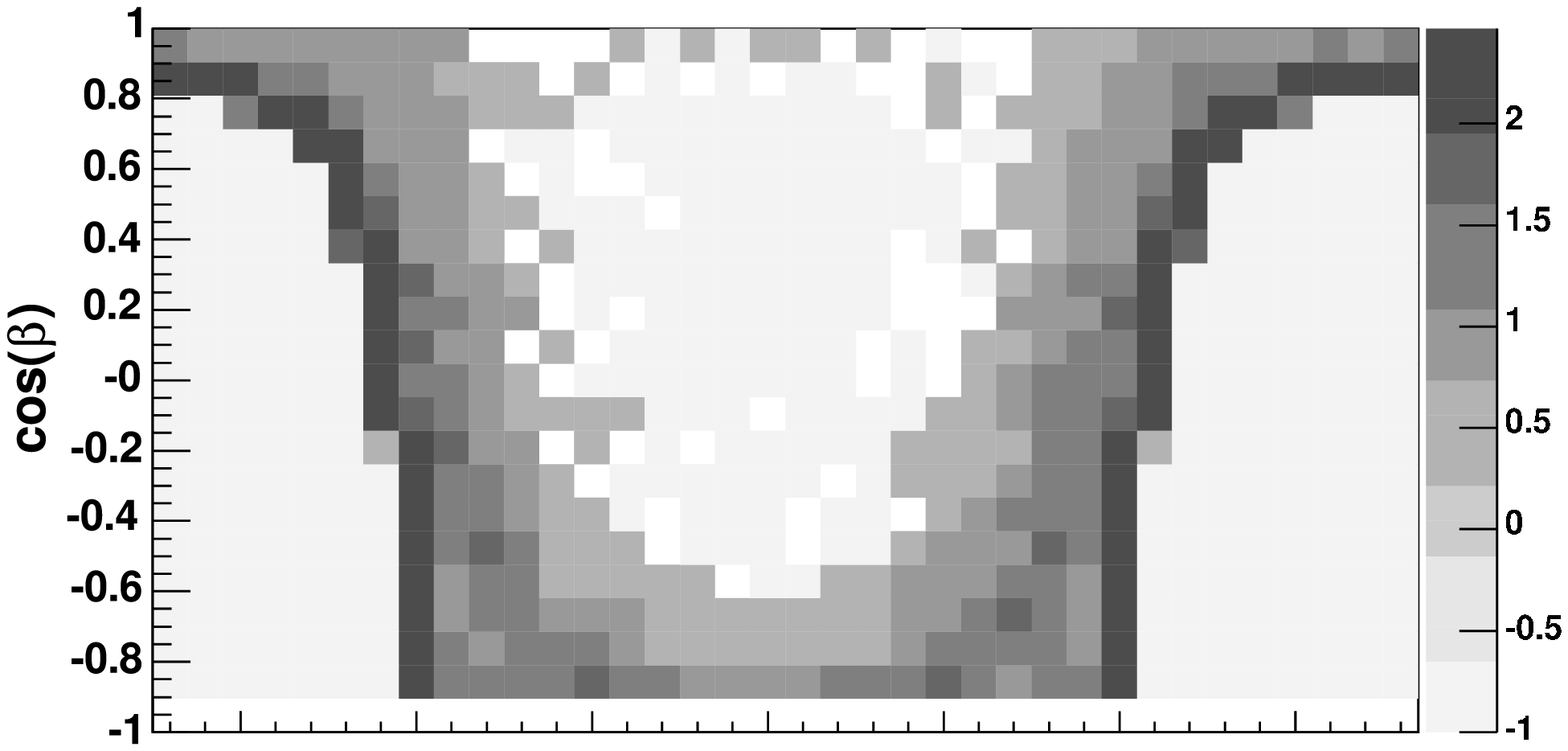}
\includegraphics{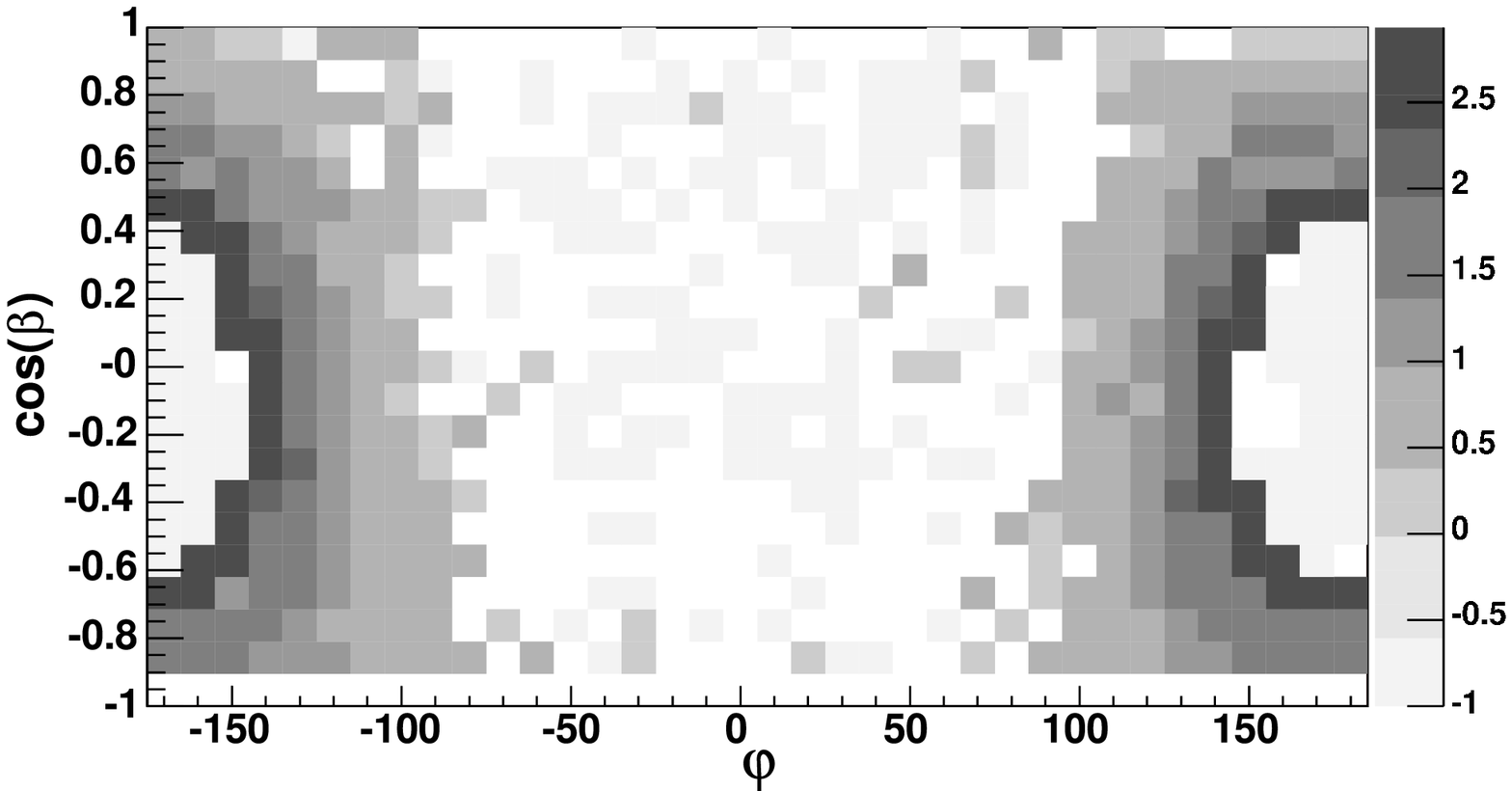}
\caption{As in Fig.~(\ref{fig5a}) but for the primary leptons with the power law 
spectrum (described in the text) which energy is also normalized to $10^6$ MeV. 
The results are averaged over 500 simulated primary leptons for the angle 
of injection $\theta = 90^o$ (a), 180 leptons for 
$120^o$ (b), and 150 leptons for $150^o$ (c).}
\label{fig5b}
\end{figure}
\subsection{Angular distribution of escaping gamma-rays}

Let us investigate the angular distribution of secondary $\gamma$-rays produced 
by secondary $e^\pm$ pairs which in turn originate in the absorption process of 
$\gamma$-rays from PWZ cascade. These $e^\pm$ pairs  
propagate in the magnetic field and anisotropic soft radiation.
As before, we consider two cases of the monoenergetic and power law injection of primary
leptons. 
In Figs.~\ref{fig5a} and \ref{fig5b} the distribution of secondary $\gamma$-rays
on the sky are shown in the case of primary leptons injected 
at specific directions defined by the angles $\theta = 90^o$ (Figs.~\ref{fig5a}a
and \ref{fig5b}a), $120^o (Figs.~\ref{fig5a}b and \ref{fig5b}b)$, and $150^o$ 
(Figs.~\ref{fig5a}c and \ref{fig5b}c). These numbers are shown for 
one injected monoenergetic lepton with energy $10^6$ MeV (Fig.~\ref{fig5a}) and 
for the power law spectrum of leptons with the power normalized to $10^6$ MeV
(Fig.~\ref{fig5b}).
The $\gamma$-rays are sorted by their escape directions, i.e. within the 
regions on the sky defined by the azimuthal directions  
($\varphi = 180^o$ corresponding to the case of the massive star in front of the pulsar) 
and by the cosine of the vertical angle $\beta$.
The basic features of the angular distribution of the secondary $\gamma$-rays 
are determined by the structure of the magnetic field in the MSWR. For the parameters
considered in this work, the magnetic field is mainly radial in the region above the
shock. For the primary leptons injected at the angle $\theta = 90^o$,
the $\gamma$-rays from cascade inside the MSWR are mainly produced 
inside the cone centred along the direction of the magnetic field lines which cross the 
direction of propagation of $\gamma$-rays produced inside the PWZ. Since the magnetic 
field lines are radial, the angles at which the secondary $\gamma$-rays escape are 
limited to relatively small part of the sphere (see Figs.~\ref{fig5a}a and \ref{fig5b}a). 
For primary leptons injected at larger angles, see e.g. $\theta = 120^o$ and $150^o$ 
(Figs.~\ref{fig5a}bc and \ref{fig5b}bc), the cone of secondary 
$\gamma$-ray production inside the MSWR becomes broader on the sky. Most 
efficient production occurs for directions which are 
tangent to the massive star limb (see Figs.~\ref{fig5a}c and \ref{fig5b}c
for the angle $\theta = 150^o$). 
In this case a significant part of secondary $\gamma$-rays fall onto the surface of the 
massive star which angular dimensions, seen from the location of the neutron star,
are equal to $26.4^o$ (note the avoidance region centred on $\varphi = 180^o$ and 
$\cos\beta = 0$ in Figs.~\ref{fig5a}c and \ref{fig5b}c).       
If the primary leptons are injected exactly towards the massive star ($\theta = 180^o$),
then the secondary $e^\pm$ leptons, which originate in the MSWR in absorption of 
secondary $\gamma$-rays from PWZ, move along the radial magnetic field. Most of the 
$\gamma$-rays produced in MSWR fall then onto the massive star. 
The secondary $e^\pm$ pairs reach finally the dipole part of the magnetic field, which 
is close to the massive star, and may produce low energy $\gamma$-rays at wide angles
in the plane of the binary system. 

The angular distribution of $\gamma$-rays produced in the MSWR in the case of 
monoenergetic and power law distributions of primary leptons is similar
since it is determined by the radial structure of the magnetic field. However, the 
numbers of produced $\gamma$-rays are a factor of a few larger in the case of 
monoenergetic injection of primary leptons. This is due to the fact that $\gamma$-rays
produced in the PWZ by monoenergetic leptons have on average higher energies
(see Figs.~\ref{fig3}ac). Therefore they transport 
more energy to the MSWR. The energies and numbers of secondary $e^\pm$ pairs,
from their absorption inside the MSWR, are larger allowing more efficient 
production of next generation of $\gamma$-rays in the MSWR.

\begin{figure}
\vspace{9.5truecm}
\includegraphics{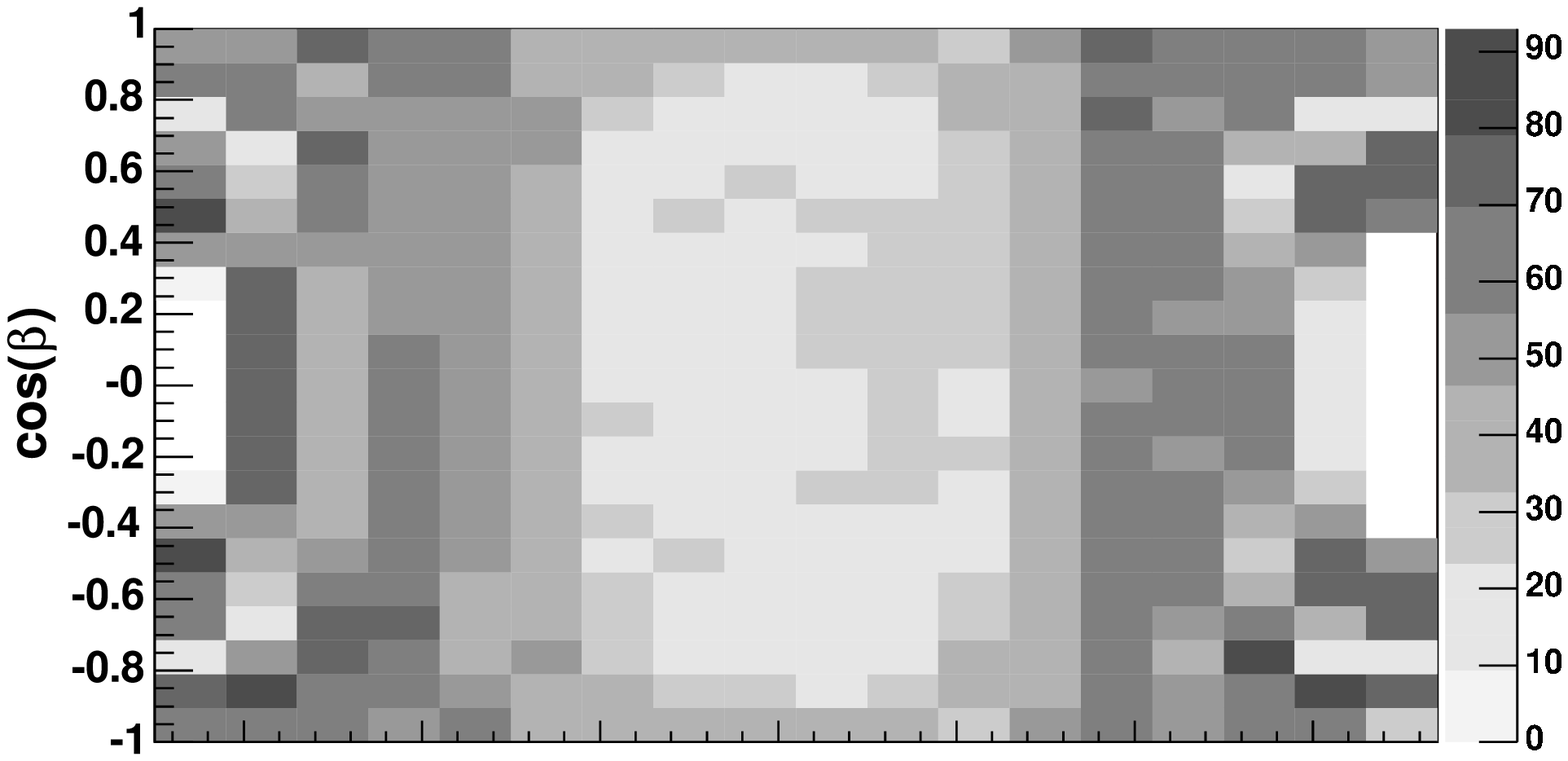}
\includegraphics{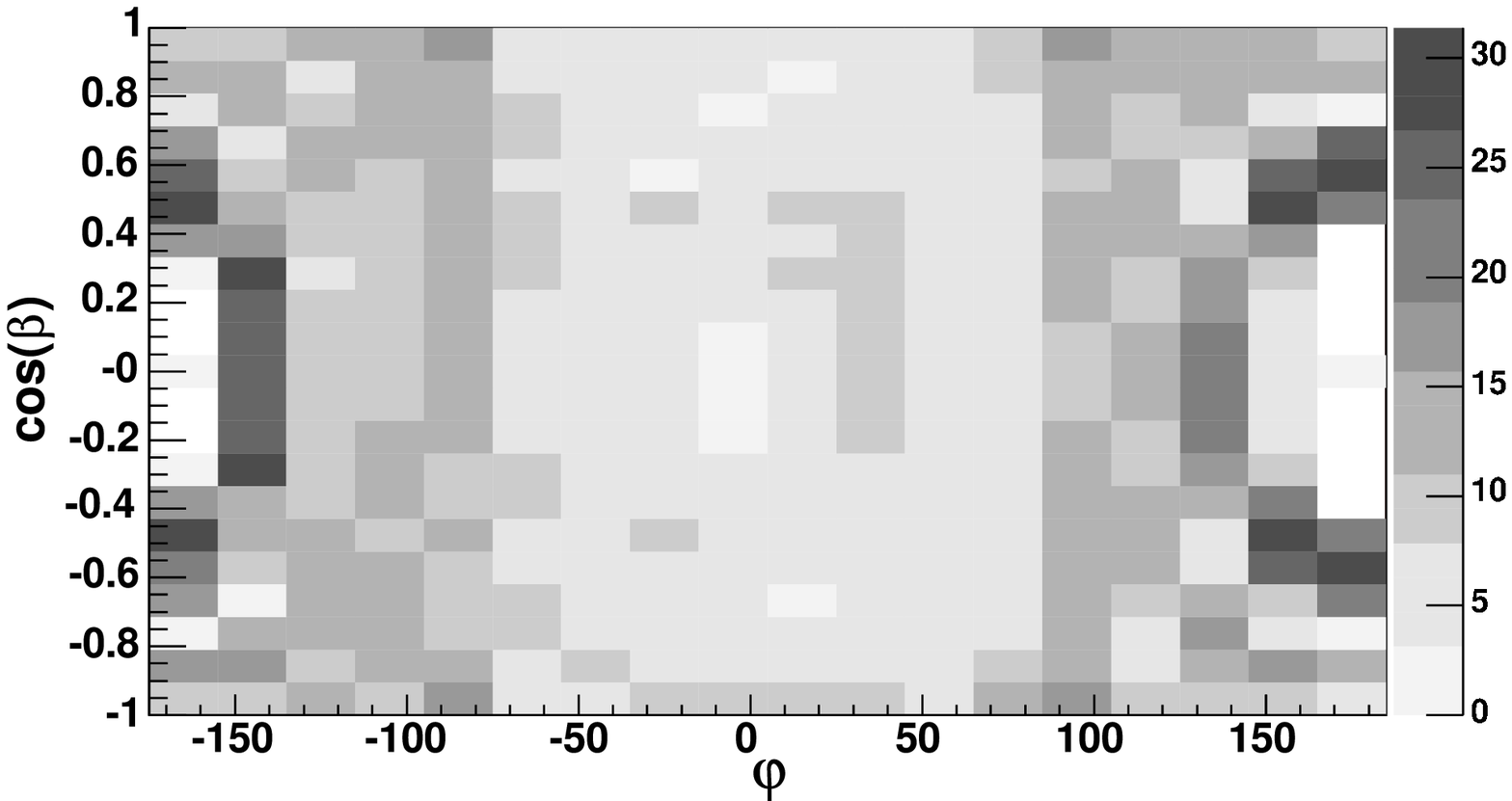}
\caption{The map with the total numbers of $\gamma$-ray photons with energies 
$> 100$ MeV (in linear grey scale), escaping from the binary at specific directions on the sky 
(defined by the azimuthal angle $\varphi$ and cosine of horizontal angle $\cos\theta$) 
for the case of isotropic injection of 
primary monoenergetic leptons with energies $10^6$ MeV (a) and for the primary leptons 
with the power law spectrum described in the text (b). $\gamma$-rays are produced in 
cascades inside the MSWR 
initiated by secondary cascade $\gamma$-rays from the PWZ. The spectra of primary 
leptons are normalized to their total energy equal to $10^6$ MeV per steradian.}
\label{fig5c}
\end{figure}

The angular distributions of $\gamma$-rays with energies $> 100$ MeV which 
escape to the observer from the MSWR for the case of isotropic injection 
of primary leptons are shown in Fig.~\ref{fig5c}. The numbers of 
these $\gamma$-rays are normalized to the power in primary lepton spectrum equal to
$10^6$ MeV sr$^{-1}$. As before, the pulsar is behind the massive star for the phase
$\varphi = 180^o$. As expected, the numbers of produced $\gamma$-rays
are the largest in directions tangent to the limb of the massive star. 
Significantly lower numbers of $\gamma$-rays emerge from the binary system in the outward
directions, i.e. in directions for which the pulsar is in front of the massive star
with respect to the observer. Note that in these directions the MSWR does not extend 
(see Fig.~1). 

\subsection{Gamma-ray light curves and spectra} 

\begin{figure}
\vspace{6.3truecm}
\includegraphics{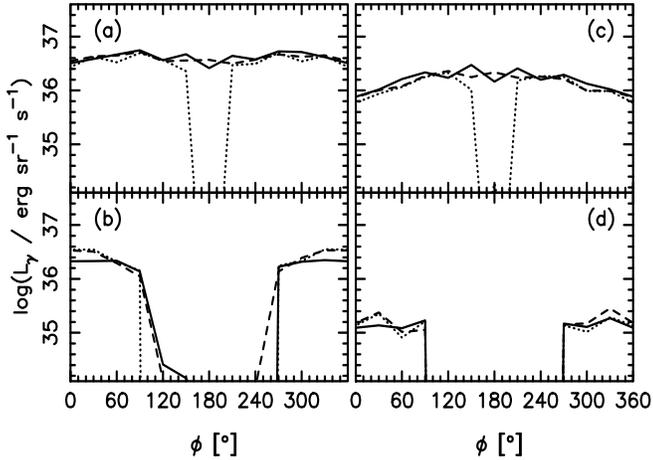}
\caption{The $\gamma$-ray light curves in two energy ranges,
$0.1 - 10$ GeV ((a) and (c)) and $10 - 10^3$ GeV ((b) and (d)),
produced in the MSWR cascade initiated by secondary $\gamma$-rays which are in turn 
produced in the cascade in the PWZ initiated by primary leptons with the monoenergetic 
(figures on the left) and the power law distributions (on the right).
$\gamma$-rays escaping from the MSWR are collected within the range of the cosine of
inclination angles of the system $0.76 < \cos i < 0.88$ (contains $i=30^o$ - 
solid curve), $0.41 < \cos i < 0.53$ ($60^o$ - dashed), and 
$-0.06 < \cos i < 0.06$ ($90^o$ - dotted).
} 
\label{fig6}
\end{figure}

The $\gamma$-ray light curves of photons produced in
the MSWR (Table 4) for two energy ranges, $0.1 - 10$ GeV and 10 GeV - 1 TeV,
are shown in Figs.~\ref{fig6}, after its normalization to the total energy loss rate 
of the considered here pulsar. The observer is located at different 
inclination angles $i$ to the plane of the binary system.
The general features of the $\gamma$-ray light curves produced in the MSWR are
similar to these ones shown above for the $\gamma$-rays produced in the PWZ (see for 
comparison Figs.~\ref{fig6}). However, the total $\gamma$-ray power is usually
lower (typically by a factor of three) than expected in the case of $\gamma$-rays 
produced in the PWZ. Only in the range of phases around $0^o$, and large inclination 
angles, the 
power of the $\gamma$-rays from the MSWR can dominate over the PWZ $\gamma$-rays
in the GeV energy range. 

The power emitted in GeV energies is quite uniform with the phase of the
pulsar provided that the inclination angle is small enough that the binary system is not
eclipsing. The power emitted at TeV energies is limited to the range of phases 
$\pm 60^o$ around the phase $0^o$ which correspond to the position of the pulsar in 
front of the massive star.

\begin{figure*}
\vspace{8.5truecm}
\includegraphics{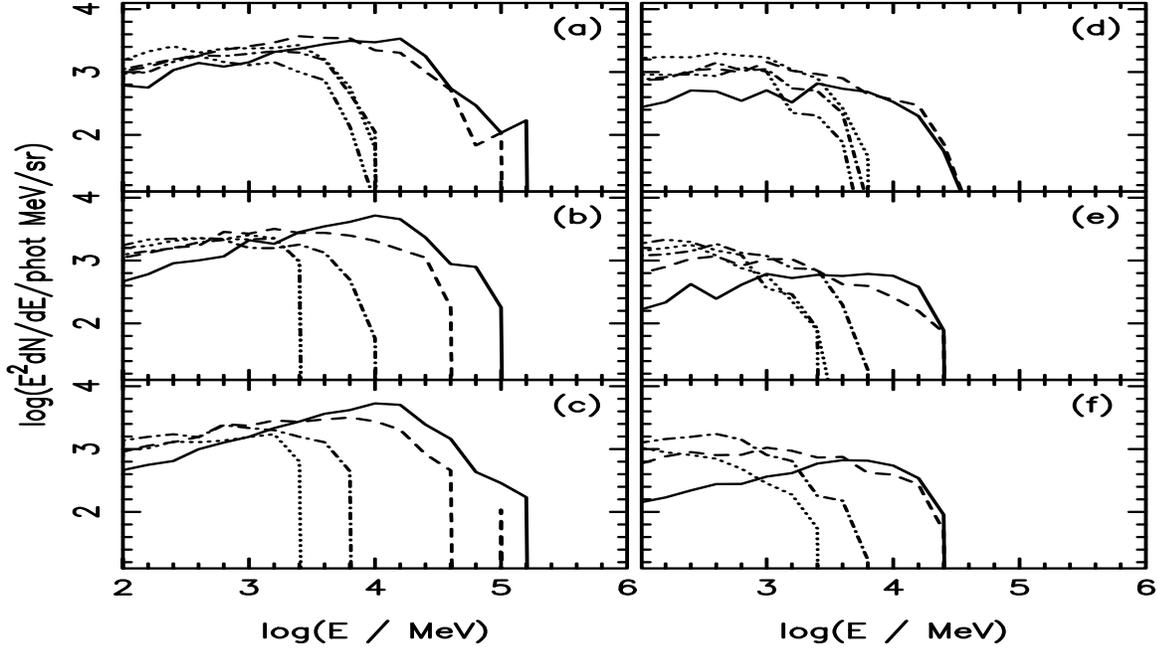}
\caption{The $\gamma$-ray spectra escaping from the MSWR for different phases of 
the pulsar with respect to the location of the observer, $\varphi = 30^o$ (solid curve) 
$60^o$ (dashed), $90^o$ (dot-dashed), $120^o$ (dotted), $180^o$ (dot-dot-dot-dashed),
and three ranges of the cosine of the inclination angle of the binary
$0.76<\cos \alpha <0.88$ (centered on $i=30^o$ - (a) and (d)), 
$0.41<\cos \alpha <0.53$ ($60^o$ - (b) and (e)), and 
$-0.06<\cos \alpha <0.06$ ($90^o$ - (c) and (f)). 
$\gamma$-rays are produced in the MSWR cascade by secondary $e^\pm$ pairs which in turn
originate in the absorption of $\gamma$-rays produced in the PWZ 
by primary leptons with the monoenergetic (figures on the left) and power law spectrum 
(on the right).} 
\label{fig7}
\end{figure*}

Figs.~\ref{fig7} show the $\gamma$-ray spectra produced in the MSWR for selected phases
and inclination angles of the binary system. The basic differences between these 
spectra and the $\gamma$-ray spectra produced in the PWZ are due to the fact that 
cascade in the MSWR is initiated by leptons for which the total 
optical depths are higher. Therefore, these 
$\gamma$-ray spectra are steeper, with the spectral index close to -2, in contrast to  
the spectral index of the spectra produced in the PWZ which are close to -1.5. 
Although, the power emitted in $\gamma$-rays from the MSWR is a factor of a few lower. It can still 
significantly contribute to the total $\gamma$-ray spectrum observed at some range 
of phases and at energies below $\sim 1$ GeV for the case of 
monoenergetic injection of primary leptons. Moreover, as already noted above, 
this GeV emission is much more uniform over the sky. At energies $> 10$ GeV
the contribution of the $\gamma$-rays from the MSWR (produced by monoenergetic primary 
leptons) becomes negligible with respect to 
the $\gamma$-rays from  the PWZ for the considered location of the shock inside 
the binary system defined by $\eta = 0.3$. The  
$\gamma$-rays produced inside the MSWR by primary leptons with the power law spectrum 
(see Figs.~\ref{fig7}def) do not contribute significantly to the $\gamma$-rays produced 
in the PWZ (Figs.~\ref{fig5}def). 
Primary leptons with the power law spectrum produce on average more secondary 
$\gamma$-rays already inside the PWZ.

\section{Dependence on the shock localization}

Since detailed parameters of the massive stellar winds are not precisely known
we investigate the dependence of escaping $\gamma$-ray spectra on the location of the 
shock inside the binary system by performing calculations for the 
same initial injection spectra of leptons but for the shock determined by the value of 
$\eta = 0.06$. This value is obtained for upper limits of the stellar wind parameters 
(see Sect.~2). For $\eta = 0.06$ the shock is located at the distance of $\rho_{\rm o} = 0.25\ R_{\rm s}$ 
from the pulsar, i.e. about two times closer than for $\eta = 0.3$, 
and the volume for the development of the cascade in the PWZ is much smaller than for the 
cascade in the MSWR.   

\subsection{Gamma-rays from PWZ}

\begin{figure}
\vspace{6.3truecm}
\includegraphics{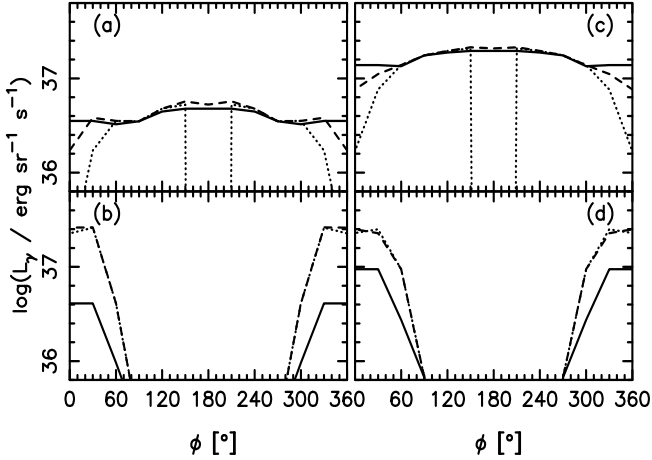}
\caption{As in Fig.~4 but for $\eta = 0.06$.} 
\label{fig8}
\end{figure}

\begin{figure*}
\vspace{8.5truecm}
\includegraphics{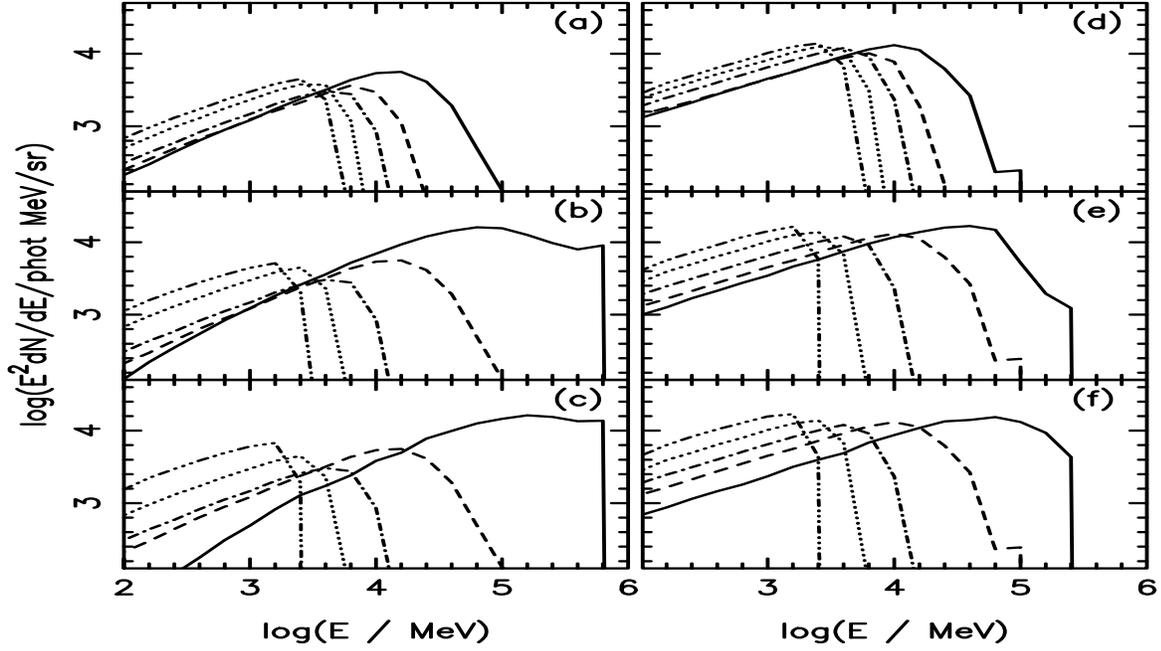}
\caption{As in Fig.~5 but for $\eta = 0.06$.} 
\label{fig9}
\end{figure*}

For leptons injected with the power law spectrum, the
$\gamma$-ray luminosities produced in the PWZ (and detected by the observer at the 
infinity) for $\eta = 0.06$ are only slightly lower than 
for $\eta = 0.3$ (see for comparison the 
$\gamma$-ray light curves and spectra in Figs.~\ref{fig8} and \ref{fig9} with
Figs.~\ref{fig4} and \ref{fig5}, respectively). 
This small differences are not surprising since the energy from
primary leptons with the power law spectrum is transfered to $\gamma$-rays 
efficiently already inside the PWZ due to large optical 
depths for leptons with energies much lower than $10^6$ MeV (see Fig.~\ref{fig2}).
However, the reduction of $\gamma$-ray fluxes is much stronger for leptons injected 
with the monoenergetic spectrum and energies $10^6$ MeV. 
The $\gamma$-ray light curves and spectra 
for the monoenergetic primary leptons and the shock defined by $\eta = 0.06$ are by a 
factor of about 3 lower than 
for $\eta = 0.3$. As shown in Fig.~\ref{fig2}c, the optical depths for leptons with 
energies $10^6$ MeV are close to unity for $\eta = 0.06$ but much larger for 
$\eta = 0.3$. Therefore, the differences in the cascades developed in the PWZ by  
high energy monoenergetic leptons are larger than reported for leptons with the power law
spectrum.

Large differences between these two discussed localizations of the shock in the 
$\gamma$-ray light curves in the energy range $10 - 10^3$ GeV
are also observed for small inclination angles $i$ 
(see Figs.~\ref{fig4}b and \ref{fig8}b). They are due to lower optical depths for 
monoenergetic leptons with energies $10^6$ MeV. Leptons injected in directions 
perpendicular to the plane of the binary 
system meet on its propagation through the PWZ smaller number of 
soft photons from the massive companion than these ones propagating towards the massive
star due to the geometrical effects. 
Moreover, the spectra of $\gamma$-rays escaping to the infinity for $\eta = 0.06$ 
show cut-offs at lower energies than in the case of the shock defined by $\eta = 0.3$. 
This effect, especially clear for the monoenergetic injection of primary leptons, is due 
to the stronger absorption of higher energy $\gamma$-rays during their propagation in the 
MSWR which volume is larger for $\eta = 0.06$ than for $\eta = 0.3$.

\subsection {Gamma-rays from MSWR}

\begin{figure}
\vspace{6.3truecm}
\includegraphics{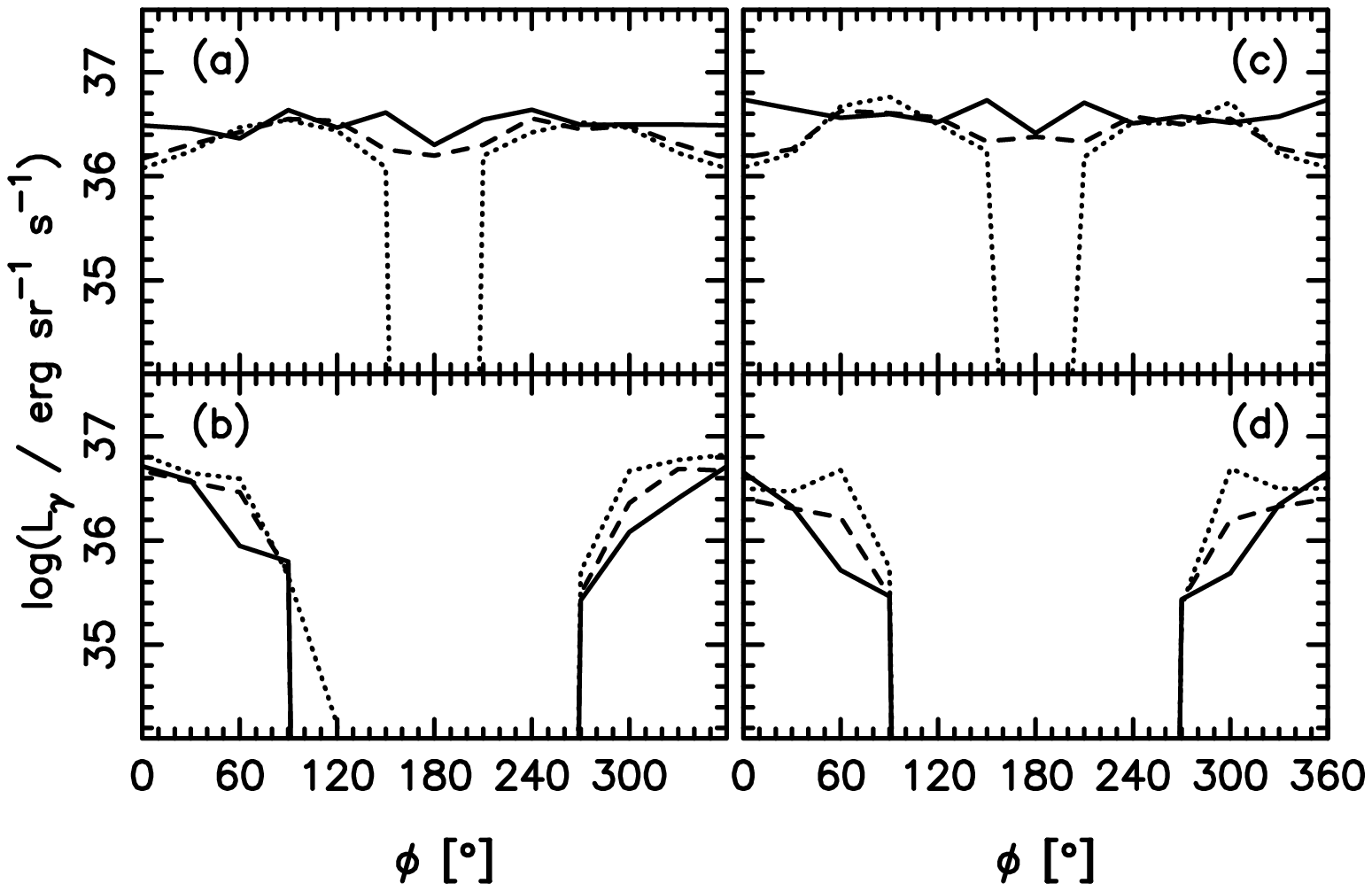}
\caption{As in Fig.~6 but for $\eta = 0.06$.} 
\label{fig10}
\end{figure}

\begin{figure*}
\vspace{8.5truecm}
\includegraphics{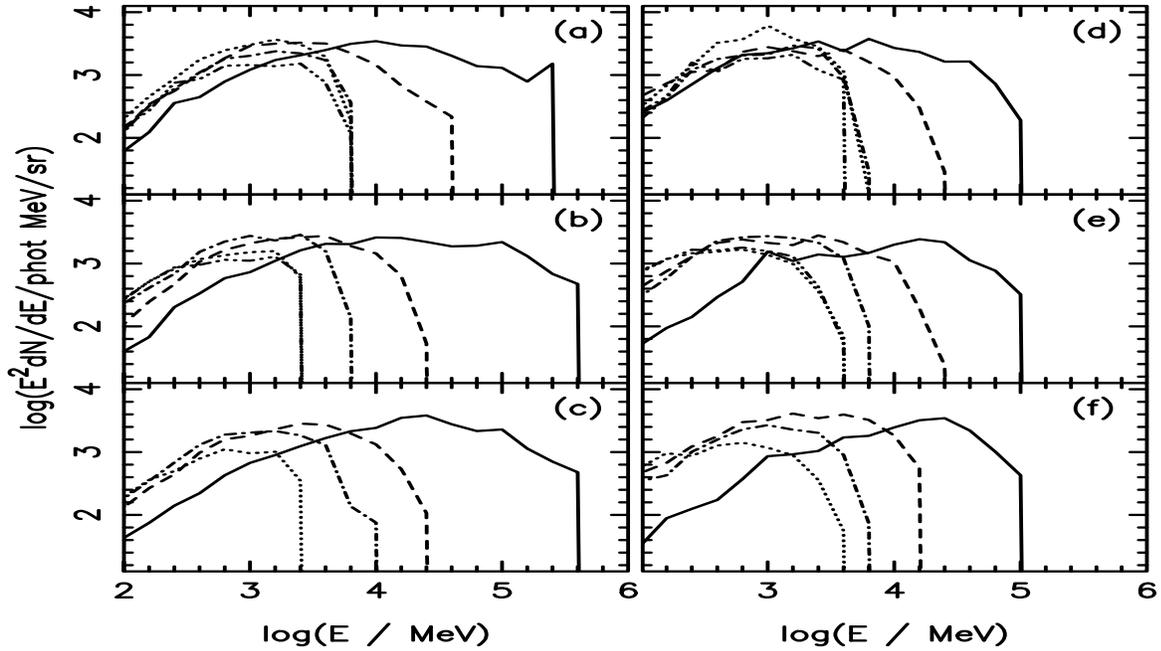}
\caption{As in Fig.~7 but for $\eta = 0.06$.} 
\label{fig11}
\end{figure*}

Basic differences in the $\gamma$-ray production in the MSWR due to
different localizations of the shock inside the massive binary (defined by $\eta = 0.3$ 
and 0.06) are related to the efficiency of the cascades occurring already inside the PWZ
since $e^\pm$ pairs, created in absorption of the PWZ $\gamma$-rays, are responsible for 
production of $\gamma$-rays in the MSWR. For localization of the shock described by 
$\eta = 0.06$, the $\gamma$-ray spectra produced in the MSWR extend to higher energies 
than for the case of $\eta = 0.3$ (see Figs.~\ref{fig7} and \ref{fig11}). This is also 
clearly seen when investigating the $\gamma$-ray light curves at energies $>10$ GeV for 
the case of the power law spectrum of primary leptons shown in Figs.~\ref{fig6}d and 
\ref{fig10}d. This behavior is in contrast to the features of the $\gamma$-ray spectra 
produced in the PWZ. 
It is clear from the Table~1b that, although the $\gamma$-ray luminosity leaving the PWZ 
(i.e. entering the MSWR) for the power law spectra of primary leptons
is comparable for $\eta = 0.3$ and 0.06 (see percentages marked by $\it shock$), the 
power transformed to the secondary $e^\pm$ pairs in the MSWR, from absorption of 
the PWZ $\gamma$-rays, differ significantly due to much larger
volume of the MSWR for $\eta = 0.06$
(see differences between percentages marked by $\it shock$ and $\it esc$). 
Therefore, secondary $e^\pm$ pairs produced in the MSWR for the shock located
closer to the pulsar (i.e. for $\eta = 0.06$) have, on average, higher energies and 
produce more energetic $\gamma$-rays in the MSWR. This is also the reason that the 
$\gamma$-ray light curves are on a slightly higher level for $\eta = 0.06$ 
(compare Figs.~\ref{fig6}cd with \ref{fig10}cd).  

For the case of the monoenergetic injection of primary leptons, the $\gamma$-ray 
luminosity originating in the PWZ is clearly lower for $\eta = 0.06$ (Table~1a). 
However, the energy 
transfered into the secondary $e^\pm$ pairs in the MSWR, as a result of absorption of 
the PWZ $\gamma$-rays, is comparable for both of the shock localizations
(note differences between percentages marked by 
$\it shock$ and $\it esc$ in Table~1a). Therefore, differences in the $\gamma$-ray light 
curves and spectra for the localizations of the shock inside the binary system defined 
by $\eta = 0.3$ and 0.06 are not so evident than in the case of the power law injection 
of primary leptons.

\section{Discussion and Conclusions}

We have analyzed in details the anisotropic IC $e^\pm$ cascade inside the massive binary system 
containing young pulsar.  We have considered the situation in which leptons are 
injected mono-directionally or isotropically from a place 
(identified with the pulsar) which differs from the center of the isotropic 
source of soft radiation (identified with the massive stellar companion).
The simplest geometrical situation has been analyzed in which a compact object is 
on a circular orbit around the massive star. As an example, the parameters of the massive 
binary in the Cyg X-3 system have been applied. We concentrate on the situation in which
a pulsar is energetic enough to create strong shock in collision with the wind of the 
massive star. As a consequence, two regions for the development of the cascade has to be
distinguished which differ in basic properties, i.e. the pulsar wind zone (PWZ) and the 
massive star wind region (MSWR). In the PWZ the cascade develops radially from the 
pulsar and therefore can be considered only in one direction. In the MSWR the cascade 
develops in complex magnetic field of the massive star. Therefore, complicated
propagation paths of secondary $e^\pm$ pairs in this magnetic field have to be taken into 
account. However, for typical values of the magnetic fields
in the winds of the massive stars and parameters of Cyg X-3 binary, 
the energy density of radiation from the massive star is larger than the energy density of the magnetic field.
Hence, the IC energy losses of leptons dominate over their synchrotron
losses and the influence of synchrotron process on the energy budget of leptons can be 
neglected in the first approximation.

For this specific binary system we have calculated the optical depths for  
leptons and $\gamma$-rays and found that they are much larger than unity 
(Fig.~\ref{fig2}). Therefore, if relativistic leptons are injected inside the binary
they should initiate efficient cascades. Such cascades are analyzed in 
the case of leptons injected by the compact object with: (a) the monoenergetic 
spectrum and energies $10^6$ MeV and, (b) the power law spectrum of the type predicted by
the calculations of the cascade processes in the inner pulsar magnetosphere 
(Hibschman \& Arons~2001).
 
The main results of our calculations are following:

\begin{enumerate}

\item Most of the energy of primary leptons is transfered to the secondary
$\gamma$-rays ($\sim 80 - 90\%$, see Table~1) for both 
the monoenergetic and the power law primary spectra, and for the shock defined by 
$\eta = 0.3$ which correspond to the case of the massive star with the wind parameters 
from the middle of the estimated range (Sect.~2). This percentage of energy does not 
depend strongly on the location
of the shock for the power law spectrum of primary leptons (compare $\eta = 0.3$ and 0.06 
in Table~1), but is significantly lower for the shock closer to the pulsar, determined by
$\eta =0.06$ in the case of monoenergetic primary leptons.
The difference between the power in $\gamma$-rays arriving to the shock (marked by 
{\it shock}) and $\gamma$-rays escaping to the observer (marked by {\it esc}) gives 
part of the power of $\gamma$-rays produced in the PWZ which is converted into the 
secondary $e^\pm$ pairs in the MSWR. These $e^\pm$ pairs can initiate additional cascade 
in the MSWR and take typically between one third and half of the $\gamma$-ray power 
escaping to the observer from the PWZ (see Table~1). 
Therefore, the contribution of $\gamma$-rays
produced in the MSWR is usually lower than $\gamma$-rays produced in the PWZ.  
However, if the monoenergetic primary leptons are injected into the binary with the shock 
located relatively close to the pulsar (e.g. for $\eta = 0.06$) the $\gamma$-ray 
luminosity produced in the MSWR 
can be even a factor of two or three larger than $\gamma$-ray luminosity from
the PWZ. This is due to lower optical depths in relatively small PWZ and so 
inefficient cascading precess initiated by high energy monoenergetic leptons.

\item The $\gamma$-ray light curves for photons escaping from the PWZ at lower energies 
(energy range $0.1 - 10$ GeV) are anticorrelated with the $\gamma$-ray light curves at 
higher energies ($> 10$ GeV) for both discussed primary spectra of leptons independently  
on localizations of the shock within the binary system (see e.g. the $\gamma$-ray light 
curves in Figs.~\ref{fig4} and \ref{fig8}).
This basic feature is also clearly seen when comparing the $\gamma$ spectra escaping
to the observer at different phases of the pulsar and inclinations of the
binary system (Figs.~\ref{fig5} and \ref{fig9}). It is easily understood since the 
intense $\gamma$-ray fluxes at energy range $0.1-10$ GeV are expected
in directions of large optical depths, when the binary system is viewed from the 
direction close to the massive star. In contrast, high level $\gamma$-ray fluxes at 
energies $>10$ GeV are expected when the compact object is in front of the massive star. 
For considered parameters of the binary system, the optical depth is high enough for 
efficient production of $\gamma$-rays above 10 GeV but is too low for efficient cascading 
and production of lower energy $\gamma$-rays.
Moreover, the $\gamma$-ray spectra extend to higher energies for the monoenergetic 
primary leptons but are less intense at lower energies than expected for the 
power law primary leptons (see Figs.~\ref{fig5} and \ref{fig9}) since the optical depths 
are lower for higher energy particles (see Figs.~\ref{fig2}).
Therefore, it is expected that detailed results in the $\gamma$-ray spectra from 
specific binaries (PSR B1259-63, SAXJ 0635+0533), for which the parameters of the 
massive star are well 
known, should allow to obtain information on the spectrum of primary
leptons accelerated by the pulsar. This problem is difficult to investigate 
directly in the case of isolated pulsars for which the surrounding soft radiation
field in the pulsar wind zone is relatively weak.  

\item The complex magnetic field in the MSWR has significant effect on the directions of
propagation of  secondary $e^\pm$ pairs and, as a consequence, on the distribution
of $\gamma$-rays on the sky. Our detailed calculations show that the angular 
distribution of $\gamma$-rays, produced in the cascade processes in the MSWR by secondary
$e^\pm$ pairs from absorption of $\gamma$-rays from PWZ, strongly depends on the 
injection direction of primary leptons (see maps in Figs.~\ref{fig5a} and \ref{fig5b}). 
For example,
if the primary particles are injected perpendicular to the plane of the binary system
then the $\gamma$-rays produced in the cascade are collimated within the cone around the 
direction of the local magnetic field lines, i.e. in the outward direction with respect to 
the massive star (Figs.~\ref{fig5a}a and \ref{fig5b}a). 
This is due to the strong radial component of the massive star
magnetic field in the main volume of the binary system (see Eq.~\ref{eq9}).
Therefore, in the case of highly anisotropic injection of 
primary particles by the pulsar, the strong $\gamma$-rays fluxes may appear in 
quite different directions.
In contrast, if the primary particles are injected in the directions not far from the
the stellar limb, this radial component of the magnetic field tends to focus the
$\gamma$-rays in directions tangent to the stellar limb. A significant part
of these $\gamma$-rays impinge onto the surface of the massive star 
(see Figs.~\ref{fig5a}c and \ref{fig5b}c).
It is expected that these $\gamma$-rays can excite observable nuclear $\gamma$-ray line 
fluxes due to their interactions with the matter on the massive star surface. This line
emission should be correlated with the high energy $\gamma$-ray emission ($> 10$ GeV) 
and anticorrelated with the low energy $\gamma$-ray emission (between $0.1 - 10$ GeV).

The distribution of $\gamma$-rays produced in the MSWR is also strongly anisotropic
for the case of isotropic injection of primary leptons by the pulsar 
(see Fig.~\ref{fig5c}).
There is a strong concentration of produced secondary $\gamma$-rays in directions
close to the limb of the massive star. The number of secondary photons produced 
by the primary leptons with the power law spectrum is by a factor of a few lower
than for monoenergetic primary leptons but general emission pattern on the sky does
not differ significantly. These features are due to the fact that leptons with lower 
energies produce $\gamma$-rays with energies above 100 MeV less efficiently.

\item The $\gamma$-ray fluxes produced in the MSWR are weakly dependent on the 
phase of the pulsar and the inclination angles of the binary system, except for the 
case of eclipsing binaries (see Figs.~\ref{fig6} and \ref{fig10}). This is due to the 
propagation effects of 
secondary $e^\pm$ pairs in the magnetic field of the MSWR which result in a significant 
isotropisation of the secondary $e^\pm$ pairs with respect to the directions of
propagation of primary leptons. Therefore, when the pulsar is in front
of the massive star (phase $\varphi = 0^o$), the main contribution to the 
escaping low energy $\gamma$-ray flux comes from the MSWR and the anticorrelation 
between the lower energy ($0.1 - 10$ GeV) and higher energy ($>10$ GeV) $\gamma$-rays 
produced in the PWZ is partially suppressed (compare Figs.~\ref{fig4} with \ref{fig6}  
and \ref{fig8} with \ref{fig10}). 
The $\gamma$-ray spectra escaping to the observer from the MSWR are flatter in the 
GeV energy range (spectral index close to -2) than the $\gamma$-ray spectra from
the PWZ (spectral index close to -1.5) (compare Figs.~\ref{fig5} with \ref{fig7} and 
\ref{fig9} with \ref{fig11}). This is due to the difference in the optical depth
for particles propagating in the PWZ and the MSWR. Since the cascade is not efficient
enough in the PWZ, the secondary $e^\pm$ pairs are not able to influence 
lower energy part of the IC $\gamma$-ray spectrum formed mainly by cooling of primary 
leptons. However, secondary $e^\pm$ pairs in the MSWR are cooled to energies low enough 
to produce a lot of secondary GeV photons and thus creating steeper spectra. 
Due to the same reasons, the $\gamma$-ray spectra from the MSWR extend usually 
to lower energies and the predicted fluxes above $\sim 10$ GeV are lower.   

\item The $\gamma$-ray spectra escaping from the MSWR depend strongly also on the 
localization of the shock inside the binary. This is caused by the change in relative
volume for the development of cascades in the PWZ and MSWR. If the shock is closer
to the pulsar then the $\gamma$-ray spectra from the MSWR extend to higher energies 
(see Figs.~\ref{fig7} and \ref{fig11}) since they are produced by on average more 
energetic secondary $e^\pm$ pairs originated in the absorption of more energetic $\gamma$-rays escaping 
from the PWZ. The $\gamma$-ray fluxes at energies $< 10$ GeV, produced in the 
MSWR, do not depend strongly on the localization of the shock. But the $\gamma$-ray fluxes 
at energies $>10$ GeV 
are on a much lower level for the case of primary leptons injected with the power law
spectrum (see Figs.~\ref{fig6} and \ref{fig10}) from the same reasons as discussed  in this item 
above.

\end{enumerate}

In summary, detailed investigation of the phase resolved IC continuous  
$\gamma$-ray spectra and the $\gamma$-ray light 
curves from the compact massive binaries, in which energetic pulsar is responsible for the
acceleration of relativistic leptons, should allow extraction of information 
on the primary spectra of leptons (monoenergetic, power law ?) and conditions for 
collisions of the pulsar and stellar winds. 
The fluxes and spectra of escaping $\gamma$-rays should be also sensitive to
the acceleration site of the primary leptons, inside the inner pulsar magnetosphere
(power law spectrum ?), close to the pulsar light cylinder (monoenergetic ?), or
during the propagation in the pulsar wind zone 
(see e.g. re-acceleration model of Contopoulos \& Kazanas~2002). 
The effects on the $\gamma$-ray spectra caused by leptons which are additionally 
accelerated in the pulsar wind zone have been recently investigated with the application 
to the PSR B1259-63/Be binary system (Sierpowska \& Bednarek~2004b), but they need 
further more detailed calculations. 

Taking into account sensitivities of the future and already operating $\gamma$-ray
telescopes in the energy range below a few hundred GeV (which are estimated on 
$\sim 3\times 10^{-13}$ erg cm$^{-2}$ s$^{-1}$ for the GLAST and $\sim 10^{-12}$  
erg cm$^{-2}$ s$^{-1}$ for the MAGIC), the type of the massive binary considered 
in this paper with the $\gamma$-ray luminosity of the order of $\sim 10^{37}$ erg 
sr$^{-1}$ s$^{-1}$ (similar to the Cyg X-3 massive binary) should be detected at any 
location within our Galaxy and also within the Magellanic Clouds.

In the forthcoming paper (Sierpowska \& Bednarek 2004a) 
we intend to apply such general model for the 
$\gamma$-ray production in specific massive binaries in which case there are evidences 
of the existence of energetic pulsars (e.g. PSR B1259-63, SAX J0635)
or compact objects with jets (see the list of sources in Table~1 in 
Mirabel \& Rodriguez~1999). Note that in the case 
of less compact binaries than considered here, the secondary 
cascade $e^\pm$ pairs arrive to the shock with 
significant energies since the optical depths for leptons are much lower. 
These pairs move then along the shock plane. Their contribution to the
total escaping $\gamma$-ray fluxes has to be also taken into account. It has not been
considered in this paper since, for the parameters of the massive star in the Cyg X-3, 
these pairs take only a relatively small part of energy of primary leptons (typically 
less than $\sim 20\%$ of the total injected energy, see Table~1). 
Therefore, relative importance of processes occurring in the PWZ, the shock 
region, and the MSWR can differ significantly for the very close binaries with the
pulsars on circular orbits (as considered here) from the binaries of the PSR 1259-63 type,
i.e. with a broad, highly eccentric orbit and a massive Be star producing non-spherical
stellar wind.

We also plan to investigate possible importance of the cascade processes inside the 
binary systems of two very massive stars which create strong shock as a result of their 
stellar wind collisions, e.g. $\gamma$-2 Vel (WC8+O7.5). Energetic $e^\pm$ pairs can 
appear inside such binaries as a result of interaction of hadrons, accelerated by the 
shock, with radiation field of the massive star.

\section*{Acknowledgments}
We thank the referee, Dr G.E. Romero, for many useful comments and suggestions.
This work is supported by the Polish KBN grants No. 5P03D02521 and 
PBZ KBN 054/P03/2001.


\begin{thebibliography}{99}

\bibitem{aa91} Aharonian, F. A., Atoyan, A.M. 1991, ApJ 381, 220
\bibitem{aa99} Aharonian, F. A., Atoyan, A.M. 1999, MNRAS, 302, 253
\bibitem{at02} Atoyan, A.M., Aye, K.-M, Chadwick, P.M. et al. 2002, A\&A 383, 864
\bibitem{asc79} Arons, J., Scharlemann, E.T. 1979, ApJ 231, 854
\bibitem{asc83} Arons, 1983, ApJ 266, 215
\bibitem{bl01} Ball, L., Dodd, J. 2001, Publ.Astron.Soc.Aust. 18, 98 
\bibitem{bk00} Ball, L., Kirk, J.G. 2000, APh 12, 335
\bibitem{bed93} Bednarek, W. 1993, A\&A 278, 307
\bibitem{bed97} Bednarek, W. 1997, A\&A  322, 523 
\bibitem{bed00} Bednarek, W. 2000, A\&A 362, 646  
\bibitem{bgkt90} Bednarek, W., Giovannelli, F., Karaku\l a, S., Tkaczyk, W. 1990, 
A\&A 236, 175 
\bibitem{br03} Benaglia, P., Romero, G.E. 2003, A\&A 399, 1121
\bibitem{br00} Beskin, V.S., Rafikov, R.R. 2000, MNRAS 313, 433
\bibitem{ba00} Bogovalov, S.V., Aharonian, F.A. 2000, MNRAS, 313, 504
\bibitem{bp04} Bosch-Ramon, V., Paredes, J.M. 2004a, A\&A 417, 1075
\bibitem{bp04b} Bosch-Ramon, V., Paredes, J.M. 2004b, in press
\bibitem{braz90} Brazier, A., et al. 1990, ApJ 350, 745 
\bibitem{chetal98} Chadwick, P.M., Dickinson, M.R., Dipper, N.A. et al. 1998, ApJ 503,
391
\bibitem{chetal99} Chadwick, P.M., Lyons, K., McComb, T.J.L. et al. 1999, ApJ 513,
161
\bibitem{car92} Carraminana, A. 1992, A\&A 264, 127
\bibitem{chr86} Cheng, K.S., Ho, C., Ruderman, M. 1986, ApJ 300, 500
\bibitem{cr91} Cheng, K.S., Ruderman, M. 1991, ApJ 373, 187
\bibitem{cher94} Cherepashchuk, A., Moffat, A. 1994, ApJ 424, 53 
\bibitem{coka02} Contopoulos, I., Kazanas, D. 2002, ApJ 566, 336
\bibitem{dh82} Daugherty, J.K., Harding, A.K. 1982, ApJ 252, 337
\bibitem{eu93} Eichler, D., Usov, V. 1993, ApJ 402, 271
\bibitem{geor01} Georganopoulos, M., Aharonian, F.A., Kirk, J.G. 2002, A\&A 388, L25
\bibitem{gw87} Girard, T., Wilson, L.A. 1987, A\&A 183, 247
\bibitem{ha03} Hall, T.A., Bond, I.H., Bradbury, S.M. et al. 2003, ApJ 583, 853
\bibitem{ha85} Hamann, W.R. 1985, A\&A 145, 443
\bibitem{han00} Hanson, M., Still, M., Fender, R., 2000, ApJ 541, 308 
\bibitem{hg90} Harding, A.K., Gaisser, T.K. 1990, ApJ 358, 561
\bibitem{hib01} Hibschman, J.A., Arons, J. 2001, ApJ 560, 871 
\bibitem{jac62} Jackson, J.D. 1962, {\it Classical Electrodynamics}, 
John Willey\& Sons, New York  
\bibitem{kaetal04} Kawachi, A., Naito, T., Patterson, J.R. et al. 2004, ApJ 607, 949
\bibitem{ken84} Kennel, C.F., Coroniti, F.V. 1984, ApJ 283, 710 
\bibitem{kbs99} Kirk, J.G., Ball, L., Skjaeraasen, O. 1999, Astro.Part.Phys 10, 31 
\bibitem{lev96} Levinson, A., Blandford, R. 1996, ApJ 456, L29
\bibitem{lip90} Lipunov, V.M. 1990, {\it Astrophysics of Neutron Stars}, Stringer-Verlag
(Berlin)
\bibitem{martre81} Maraschi, L., Treves, A. 1981, MNRAS 194, 1  
\bibitem{mel95} Melatos, A., Johnston, S., Melrose, D.B. 1995, MNRAS 275, 381  
\bibitem{mr99} Mirabel, I.,F., Rodriquez, L.F. 1999, ARAA 37, 409
\bibitem{mori97} Mori, M., Bertsch, D.L., Dingus, B.L. et al. 1997, ApJ 476, 842 
\bibitem{mkt93} Moskalenko, I., Karaku\l a, S., Tkaczyk, W. 1993, MNRAS 260, 681 
\bibitem{mur03} Murata, K., Tamaki, H., Maki, H., Shibazaki, N. 2003, PASJ 55, 467
\bibitem{nt87} Nomoto, K., Tsuruta, S. 1987, ApJ 312, 711
\bibitem{or04} Orellana, M., Romero, G.E. 2004, in Proc. The multiwavelength 
approach to unidentified gamma-ray sources", Eds. K. S. Cheng \& G.E. Romero, 
Kluwer Academic Publisher (Astrophysics and Space Sciences Journal), in press 
\bibitem{pet al00} Paredes, J.M., Marti, J., Ribo, M., Massi, M. 2000, Science 288, 2340
\bibitem{ps87} Protheroe, R.J., Stanev, T. 1987, ApJ 322, 838
\bibitem{rom01} Romero, G.E., Kaufman Bernado, M.M., Combi, J.A., Torres, D.F., 
2001, A\&A 376, 599 
\bibitem{rom02} Romero, G.E., Kaufman Bernado, M.M., Mirabel, I.F. 2002, 
A\&A 393, 61 
\bibitem{rom03} Romero, G.E., Torres, D.F., Kaufman Bernado, M.M., Mirabel, I.F. 2003, 
A\&A 410, L1 
\bibitem{rs75} Ruderman, M.A., Sutherland, P.G. 1975, ApJ 196, 51
\bibitem{sch04} Schlenker, S. for the HESS collab. 2004, in Proc. Int.Symp.High Energy
Gamma-Ray Astronomy (Heidelberg, Germany), AIP, submitted 
\bibitem{sie04} Sierpowska, A. 2004, PhD thesis, in preparation
\bibitem{sb04a} Sierpowska, A., Bednarek, W. 2004a, in preparation
\bibitem{sb04b} Sierpowska, A., Bednarek, W. 2004b, in Proc. International School of 
Cosmic Ray Astrophysics - 13th Course, eds. M.M. Shapiro, T. Stanev \& J.P. Wefel 
(World Scientific), p.95
\bibitem{stark03} Stark, M., Saia, M. 2003, ApJ 587, L101
\bibitem{tak94} Tavani, M., Arons, A., Kaspi, V.M. 1994, ApJ 433, L37
\bibitem{ta97} Tavani, M., Arons, J. 1997, ApJ 477, 439 
\bibitem{tetal95} Thompson, D.J., Bertsch, D.L., Dingus, B.L. et al. 1995, ApJS 101, 259
\bibitem{um92} Usov, V.V., Melrose, D.B. 1992, ApJ 395, 575
\bibitem{vk02} van Kerkwijk, M.H., Geballe, T.R., King, D.L. et al. 2002, A\&A 314, 521
\bibitem{ves82} Vestrand, W., Eichler, D. 1982, ApJ 261, 251  
\bibitem{vsm97} Vestrand, W.T., Sreekumar, P., Mori, M. 1997, ApJ 483, L49
\bibitem{week92} Weekes, T.C. 1992, Sp.Sci.Rev. 59, 314

\end{thebibliography}
\end{document}